\documentclass[aps,prx,twocolumn,superscriptaddress]{revtex4-2}

\usepackage[utf8]{inputenc}
\usepackage{amsmath,amssymb}
\usepackage{graphicx}
\usepackage[colorlinks=true, linkcolor=black, citecolor=black, urlcolor=black]{hyperref}
\usepackage{enumerate}
\usepackage{enumitem}

\begin{document}

\title{A nonlinear hydrodynamic theory of
  ultrafast laser self-organization}

\author{J.P. Colombier}
\email{jean.philippe.colombier@univ-st-etienne.fr}
\author{E. Gandon}
\author{Q. Fornasiero}
\author{E. Brandao}
\affiliation{UJM Saint-Etienne, CNRS, Laboratoire Hubert Curien UMR 5516,
  Saint-Etienne, F-42023, France}

\begin{abstract}
Ultrashort laser pulses can induce order out of chaos.
A femtosecond pulse melts a metal surface for about a hundred
picoseconds; under photoexcitation with no net polarization axis
to imprint an orientation, the resolidifying film nonetheless
organizes, pulse after pulse, into regular nanoscale patterns.
Here we derive from first principles the nonlinear hydrodynamics
responsible for this self-organization, end to end, and show that
it accounts for the full observed phase diagram of morphologies
from a single set of measurable laser and material parameters.
Starting from the Navier--Stokes equations in the lubrication
limit, the molten film obeys a damped Kuramoto--Sivashinsky
equation that reduces to a Swift--Hohenberg formalism near
threshold, with every coefficient tied to the laser fluence, the
melt thickness, the optical skin depth, and the thermocapillary
response rather than adjustable.
The ratio of melt depth to optical skin depth sets the pattern
wavelength, while the inter-pulse delay, through post-pulse
damping, gates the route from flat surface to ordered pattern to
chaos.
We validate the resulting phase diagram (hexagonal cavity arrays,
bistable mixed states, labyrinths) numerically and confront it
with pulse-resolved experiments on two metals, Ni and Fe$_3$Cr,
which develop distinct morphologies. The agreement
between theory and experiment establishes a quantitative link
between the microscopic laser--material response and the
macroscopic pattern selected after many pulses.
We further show that this order is self-limiting: the same
corrugation that the instability builds scatters the modes that
build it, and simulations with every scale fixed by material
parameters locate the resulting breakdown of coherence near a
roughness of $30\,\mathrm{nm}$ after a few tens of pulses, a
finite process window that we confirm directly by pulse-by-pulse
measurement of orientational order on the experimental surfaces;
at stronger coupling, disorder does not destroy the pattern, it
saturates and pins it instead. This establishes a closed,
predictive description of laser-driven self-organization from the
single-pulse melt dynamics to the arrest of nanoscale order.
\end{abstract}

\maketitle


A system driven repeatedly out of equilibrium does more than
dissipate the work pumped into it: trajectories that channel part
of that work into structure, rather than into heat alone, are
statistically favored, and history accumulates through memory of
what came before~\cite{england2015dissipative}.
This dissipative logic now spans scales that have little else in
common, from self-assembling nanoparticles to living
cells~\cite{ilday2017rich,makey2020universality}, from
reconfigurable colloidal lasers~\cite{trivedi2022self} to
asymmetric soap films~\cite{kaul2024realizing}: driven far enough
from equilibrium, matter does not merely relax toward a
featureless state, it organizes.
Laser-irradiated metal surfaces make this idea testable.
Laser-induced periodic surface structures ordinarily inherit
their orientation from the incident polarization and their
periodicity from interference with a wave scattered or
plasmonically enhanced at the
surface~\cite{bonse2020maxwell,georges2025exploring,georges2026isotropic}, so the
pattern is normally dictated by the light as much as by the
matter. Surface disorder alone already breaks translational
symmetry in the absorbed intensity: rough or nanostructured metal
surfaces support transient near-field ``hot spots'' where the local
intensity can exceed the incident one by orders of
magnitude~\cite{stockman2000}, seeding the inhomogeneous absorption
that the hydrodynamic instability below amplifies.
Strip that geometry away, and order should vanish if it was never
more than a projection of the incident polarization.
It does not: under isotropic photoexcitation, obtained here with
pairs of oppositely circularly polarized femtosecond pulses that
carry no net polarization axis, a molten metal film still
organizes, pulse after pulse, into regular nanoscale patterns,
with no directional template left to inherit from.

What emerges is not contained in the physics of one pulse.
A femtosecond pulse deposits its energy within an optical skin
depth of a few tens of nanometers, over a duration of a few
hundred femtoseconds, then leaves the surface molten for about a
hundred picoseconds before
resolidification~\cite{sugioka2014ultrafast,vorobyev2013direct,stoian2020advances}:
purely local, through-depth scales, well below the
micrometer-scale footprint of the beam.
Nothing in that local process alone sets a few-hundred-nanometer
periodicity, or a corrugation that persists once frozen into the
resolidified
surface~\cite{varlamova2006self,Bonse2012Jul,Zhang2015Nov,gnilitskyi2017high}.
This is where the logic above becomes concrete hydrodynamics:
thin liquid films, with no imposed pattern, rupture their free
surface into ordered arrays of drops or ridges, quantitatively
captured by the same lubrication-theory framework developed
below~\cite{becker2003complex}.
Lateral transport within the melt and memory accumulated over
tens of pulses close the remaining gap: each pulse leaves a
topographical fingerprint that conditions the response of the
next, producing a history-dependent feedback
loop~\cite{banna2025photonic} that drives the surface through
deterministic chaos~\cite{brandao2023learning}.

The Kuramoto--Sivashinsky (KS) equation is the natural setting for
this kind of history-dependent instability.
Developed originally for flame fronts and thin-film
flows~\cite{kuramoto1978diffusion,michelson1977nonlinear,kuramoto2003chemical},
it has become a minimal model for spatiotemporal complexity in
driven dissipative systems~\cite{manneville1990dissipative}: a
linear instability amplifies long-wavelength modes, a stabilizing
term damps short-wavelength fluctuations, and a nonlinear coupling
redistributes energy between them.
The balance between these terms places KS at the boundary between
coherence and spatiotemporal
chaos~\cite{aranson2002world,round2025coherent}, and connects
infinite-dimensional field dynamics to the low-dimensional
attractors of nonlinear dynamical
systems~\cite{hyman1986bridge}: coherent structures persist inside
globally chaotic states, exactly the coexistence this paper needs
to explain.

Generalizations developed in thin-film hydrodynamics and
reaction--diffusion
systems~\cite{bestehorn2006,nepomnyashchy2010,oron1997} add
spatial couplings, memory terms, or dispersive kernels to capture
this kind of nonlocality.
The Swift--Hohenberg equation (SH), a close relative of KS that
embeds such effects directly in its operator, has proven central
for pattern selection in convective and optical
systems~\cite{swift1977,lega1994swift,bodenschatz2000recent,knobloch2015spatial,brandao2023learning}.

Damped KS-type equations recur wherever a destabilizing flux
competes with a stabilizing one: curvature-dependent sputtering or
mass redistribution against surface diffusion in ion-bombarded
surfaces~\cite{bradley1988theory,facsko1999formation,castro2005self,bradley2020theory,munoz2014self},
or interfacial thermal gradients against capillarity in the
dewetting of nanosecond-pulsed liquid
films~\cite{trice2008novel,krishna2011pulsed,kondic2020liquid}.
For LIPSS~\cite{bonse2020maxwell,bonse2024probing}, Emel'yanov and
co-workers~\cite{emel2009kuramoto,emel2011kuramoto,emel20133d} and
Reif's group~\cite{varlamova2011laser,varlamova2014genesis} used
the same KS-type dynamics, driven by defects, photoexcited
electrons, or an underdetermined balance of erosion and diffusion,
to capture wavelength selection, defect dynamics, and polarization
sensitivity.
Reif and co-workers' intuition that a Kuramoto--Sivashinsky-type
equation should govern LIPSS formation was, in hindsight, essentially
right; missing was the condition under which the isotropic reduction
actually holds, the absence of any polarization-imprinted anisotropy,
which the isotropic irradiation used below supplies and the
derivation makes explicit.
What these models postulate, this paper derives: the KS
coefficients follow directly from the thin-film equations below,
and the isotropic $|\nabla u|^2$ nonlinearity is restricted to the
isotropic forcing for which it actually
holds~\cite{Fraggelakis2018Dec,abou2020sub}, rather than applied to
LIPSS in general, including the polarized geometries where the
driving is anisotropic~\cite{rudenko2019self}.

Recent experiments confirm that nonlinear models capture this
richness, with three refinements~\cite{nakhoul2024beyond}: the
transient molten-layer thickness, set by fluence, pulse duration,
and repetition rate, filters which modes can
grow~\cite{tsibidis2012dynamics,tsibidis2015ripples}; polarization imprints anisotropy,
orienting and selecting surface
modes~\cite{zhang2020laser,perrakis2024impact}; and with increasing
pulse number, competing instabilities and feedback drive
progressively complex
morphologies~\cite{young1984laser,oktem2013nonlinear,Nakhoul2022Jul,yavuz2025controlling}.
Pattern formation, in other words, emerges from nonlinear feedback
across pulses: each one modifies the morphology the next inherits,
building up the memory-driven dynamics formalized in
Sec.~\ref{subsec:damping}~\cite{banna2025photonic}.

That lateral transport is not a minor correction.
Long-range order surviving beyond the thermal diffusion length,
the optical wavelength, and even the beam waist can only come from
nonlocal coupling, here carried by capillary flows,
thermocapillary stresses, and recoil-driven mass redistribution
linking distant regions of the molten
layer~\cite{hutt2007generalization}.
The surface responds not just to local irradiation conditions but
to its own evolving global context.

A first-principles treatment of nanoscale electron--phonon
hydrodynamics this far from equilibrium is still out of
reach~\cite{rethfeld2017modelling,rudenko2023light}: the
stochastic Navier--Stokes equations enter a regime where continuum
assumptions break down~\cite{kavokine2021fluids},
electron--phonon coupling becomes time-dependent and ill-defined
at high lattice
temperatures~\cite{anisimov1974electron,medvedev2020electron}, and
interfacial thermal transport departs from bulk
behavior~\cite{herrero2019shear}, while simulations struggle to
resolve sharp gradients, stochastic fluctuations, and nonlinear
dissipation together.
Reduced models, reaction--diffusion systems, SH, KS-type
formulations, sidestep that intractability and isolate the
mechanisms that actually select the
pattern~\cite{cross1993pattern,karma1996phase,kuramoto2003chemical}.
Our goal is to make that reduction quantitative rather than
qualitative: to carry the derivation from the Navier--Stokes
equations to the final amplitude equations without leaving a
coefficient to be fitted, so that fluence, melt thickness, skin
depth, inter-pulse delay, and thermocapillary coefficient each have
a fixed, signed effect on the instability, the selected
wavelength, and the pattern symmetry.
That end-to-end derivation is what makes the model predictive
rather than descriptive.

A complementary mechanism limits the long-range order the
KS--SH hierarchy produces: the corrugation deposited by successive
pulses is as much the product of the instability as it is a
source of disorder that scatters the same modes that build the
pattern.
This feedback between growing roughness and wave scattering points
to a disorder-driven breakdown of long-range
order~\cite{anderson1958absence,abrahams1979scaling}: once the
elastic mean free path $\ell_s$ of the critical hydrodynamic mode
becomes comparable to its wavelength ($k_m\ell_s \lesssim 1$, the
Ioffe--Regel condition~\cite{ioffe1960non,wiersma2013disordered}),
coherent pattern formation should degrade.
We construct an estimate of this scattering length from the same
parameters that enter the KS and SH coefficients, show that the
resulting coherence-degradation threshold matches the experimental
conditions explored here, and test it directly by measuring
orientational coherence length pulse by pulse on real surface
topographies.

The paper follows this arc, from isotropic excitation to order to
its own breakdown.
Section~\ref{sec:experimental} presents double-pulse experiments on
Ni and FeCr (001) single crystals under isotropic,
polarization-controlled irradiation.
Section~\ref{sec:nonlinear} carries the derivation from thin-film
hydrodynamics to the damped KS equation and its SH reduction,
every coefficient tied to laser and material parameters.
Section~\ref{sec:numerical} validates the resulting phase diagram
numerically and resolves the two-stage KS$\to$SH dynamics.
Section~\ref{sec:disorder} shows how the corrugation the
instability generates ultimately limits its own coherence, closing
the loop between the underlying instability, structural inheritance,
and disorder-induced scattering, and confirms the resulting finite
process window with a pulse-by-pulse measurement on real surfaces.

\section{Experimental observations}
\label{sec:experimental}

\begin{figure*}[t]
  \centering
  \includegraphics[width=\linewidth]{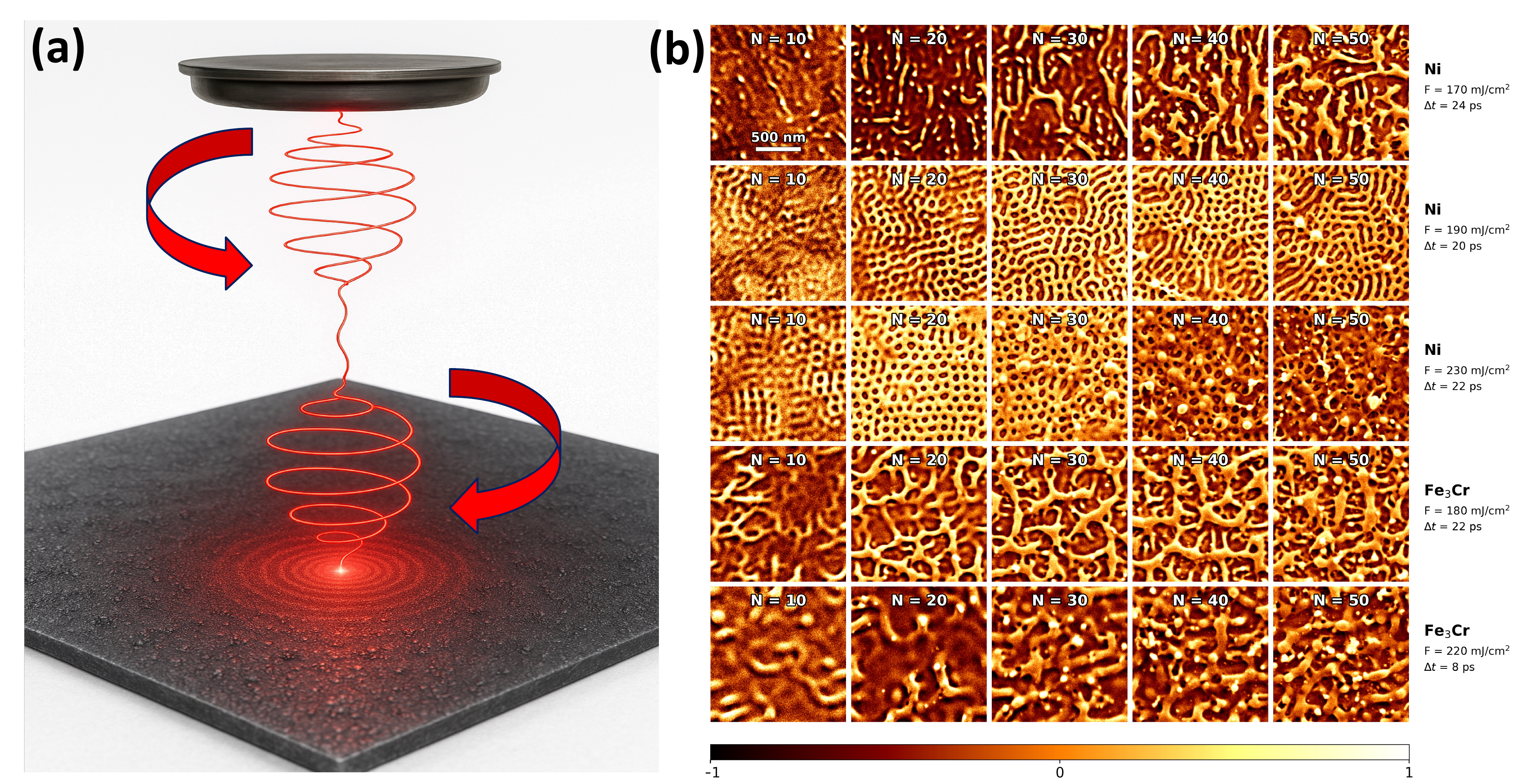}
  \caption{%
    (a)~Concept: a pair of collinear, oppositely circularly
    polarized femtosecond pulses separated by a delay $\Delta t$
    provides isotropic photoexcitation of the surface.
    (b)~Normalized surface topography versus the number of
    double pulses $N$ (columns, $N = 10$--$50$) for five irradiation
    conditions (rows): Ni at $F = 170$, $190$, and
    $230\,\mathrm{mJ\,cm^{-2}}$ ($\Delta t = 24$, $20$, and
    $22\,\mathrm{ps}$), and Fe$_3$Cr at $F = 180$ and
    $220\,\mathrm{mJ\,cm^{-2}}$ ($\Delta t = 22$ and
    $8\,\mathrm{ps}$).
    Scale bar: 500~nm.%
  }
  \label{fig:exp}
\end{figure*}

\subsection*{Isotropic photoexcitation}

A femtosecond Pharos laser system (Light Conversion;
$\lambda = 1030\,\mathrm{nm}$, pulse duration $\sim\!200\,\mathrm{fs}$,
repetition rate $50\,\mathrm{kHz}$) was employed for double-pulse
irradiation.
The average output power, tunable between $400\,\mathrm{mW}$ and
$10\,\mathrm{W}$, was initially set via software control then finely
adjusted using a variable attenuator consisting of a half-wave plate
(HWP) and Brewster-angle thin-film polarizers (TFPs), ensuring stable
linear polarization at the interferometer entrance.

The beam was split equally by a 50/50 non-polarizing beam splitter
into a modified Mach--Zehnder interferometer.
One path was fixed while the second passed through a motorized delay
line, allowing precise tuning of the inter-pulse delay $\Delta t$
from 0 to $100\,\mathrm{ps}$ with picosecond resolution.
Each interferometer arm included an independent HWP--TFP pair for
individual pulse energy control.
Circular polarization was introduced by a quarter-wave plate
($\lambda/4$): in the first arm its fast axis was set to $+45^\circ$
(left-hand circular polarization, LCP), while in the second arm it was
set to $-45^\circ$ (right-hand circular polarization, RCP).
Polarization purity was monitored with a polarimeter, confirming a
degree of circular polarization above 98\% in both arms.

The two beams were recombined at a second beam splitter and focused
onto the sample using a $250\,\mathrm{mm}$ fused-silica lens,
producing a Gaussian spot with $1/e^2$ radius of $39.2\,\mu\mathrm{m}$
confirmed by the $d^2$-method prior to each run.
Laser pulses were incident normally on the samples to ensure
consistent absorption.
This setup generated two temporally delayed, collinear femtosecond
pulses with opposite circular polarizations, an ideal condition for
studying hydrodynamic instabilities free from polarization-driven
anisotropic forcing.

\subsection*{Material preparation}

Two (001) single-crystal metals were investigated, nickel and
iron--chromium (Fe$_3$Cr, hereafter FeCr), prepared and
characterized following the same protocol throughout
(Ref.~\cite{abou2020sub} gives further detail on the nickel
samples). The FeCr crystal was grown by directional solidification,
oriented along the (001) crystallographic axis, and sectioned into
$10\times10\times10\,\mathrm{mm}^3$ specimens using a precision wire
saw; chemical analysis confirmed a composition of approximately
$74\,\mathrm{wt.\%}$ iron, $25\,\mathrm{wt.\%}$ chromium, and
$1\,\mathrm{wt.\%}$ molybdenum.

To ensure uniformity and reproducibility in laser--surface interaction,
every sample, of either metal, underwent the same multistep surface
preparation.
Mechanical polishing was performed using a Buehler Automet~250 system,
progressing through SiC abrasive papers (P180 to P2400 grit) followed
by diamond suspension polishing ($3\,\mu\mathrm{m}$ and
$1\,\mu\mathrm{m}$), removing mechanical scratches and reducing
residual topography.
This sequence yielded a mirror-finish surface with mean roughness
$R_a < 5\,\mathrm{nm}$, verified by atomic force microscopy (AFM)
over a $5\times5\,\mu\mathrm{m}^2$ scan area.

\subsection*{Laser-induced self-assembly and pattern formation}

Figure~\ref{fig:exp}(b) tracks the surface topography of the two
metals as the number of double pulses grows from $N = 10$ to $50$,
for five combinations of fluence and inter-pulse delay.
Each series is a time-lapse record of the instability.

On nickel at $230\,\mathrm{mJ\,cm^{-2}}$, a shallow, disordered
corrugation at $N = 10$ organizes into a hexagonal-like array of
circular nanocavities whose order peaks near $N \approx 20$--$30$
and then degrades by $N = 40$--$50$, so that the finite ordering
window predicted below is already visible in the raw data.
At $190\,\mathrm{mJ\,cm^{-2}}$ the cavities nucleate but
progressively elongate and interconnect, drifting toward a
labyrinth of channels; at $170\,\mathrm{mJ\,cm^{-2}}$, closer to
onset, sparse ligaments emerge and coarsen slowly without reaching
a dense pattern within $50$ pulses.
Iron--chromium behaves differently.
At both $180$ and $220\,\mathrm{mJ\,cm^{-2}}$ the surface develops
labyrinthine ligament networks from the earliest stages, which
thicken and coarsen with $N$ but never lock into a periodic cavity
lattice; at the short delay ($8\,\mathrm{ps}$), soft rounded bumps
precede the network.

Two robust facts emerge from these series.
First, the pattern scale is set early, barely evolves with $N$, and
is distinctly finer on Ni than on FeCr: wavelength selection precedes
symmetry selection. That six-fold order should emerge at all under
strictly isotropic photoexcitation, with no polarization axis left to
select it, is a striking case of spontaneous symmetry breaking.
Second, the degree of order is non-monotonic in $N$ under optimal
conditions, building, peaking and then degrading, while the morphology
class itself (cavity lattice versus labyrinth) is fixed by the
material and by $(F, \Delta t)$.

The inter-pulse delay and the fluence retain their role of fine
controls: tens of $\mathrm{mJ\,cm^{-2}}$ or a few picoseconds
separate a cavity lattice from a labyrinth on the same metal.
This sensitivity is a direct manifestation of morphological
memory: the delay controls how much of the topography inherited
from pulse $N$ survives to seed pulse $N+1$, and therefore which
instability mode is selectively amplified.
The remainder of the paper builds the theory that accounts for
these observations: the selected scale, the morphology classes,
the material contrast, and the rise and fall of order with $N$.

\section{Nonlinear surface hydrodynamics}
\label{sec:nonlinear}

This section constructs the reduced model underlying the rest of the
paper.
A guiding distinction is maintained throughout: thermocapillary and
capillary flows redistribute liquid and heat \emph{laterally} and
therefore conserve the total amount of liquid, whereas phase
transformations (solidification at the lower interface, evaporation
at the free surface) create or remove liquid \emph{locally} and
therefore do not.
Keeping the two contributions separate leads to an exact linear
theory with spatially varying coefficients, which takes a
Swift--Hohenberg form without approximation; the
constant-coefficient KS and SH models used in
the remainder of the paper are then recovered in a controlled local
limit.
Each step is carried out explicitly so that the physical origin of
every coefficient is transparent.

\subsection{Physical situation and thin-film equation}
\label{subsec:thinfilm}

\subsubsection*{Energy stored in the molten film}

\begin{figure}[!t]
  \centering
  \includegraphics[width=\columnwidth]{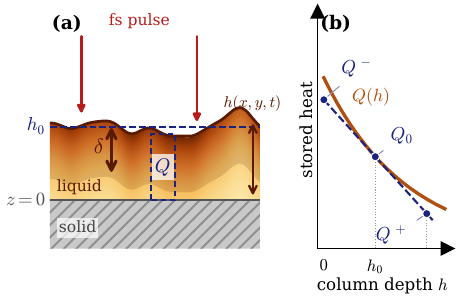}
  \caption{%
    Model geometry and notation.
    (a)~Cross-section of the molten film: solid/liquid interface
    fixed at $z=0$, local melt depth $h(x,y,t)$, unperturbed
    reference thickness $h_0$, laser energy absorbed within the skin
    depth $\delta$ and stored as excess heat $Q(x,y,h)$ in each
    liquid column.
    (b)~Stored heat $Q$ versus column depth $h$. $Q^\pm$: linear
    extrapolations of $Q(h)$ about $h_0$ to $h=0$ and $2h_0$, the two
    depths entering the perturbation expansion of
    Sec.~\ref{subsec:master}; $Q^-$ sets the thermocapillary
    anti-diffusion (growth and wavelength selection), $Q^+$ the
    drift of the perturbations.
  }
  \label{fig:schema}
\end{figure}

We consider the hydrodynamic evolution of the thin metal layer
molten by a femtosecond pulse; the geometry and notation are
summarized in Fig.~\ref{fig:schema}.
On the timescale of the flow, the laser interacts quasi-instantaneously
with matter: energy is transferred to the metal as heat in a
nonuniform, highly localized
manner~\cite{anisimov1974electron,hohlfeld2000electron,rethfeld2017modelling}.
In the non-ablative regime considered here, most of this energy is
used to liquefy the metal down to a local depth $h(x,y,t)$, where
$(x,y)$ are coordinates in the plane perpendicular to the
irradiation, and the remainder is stored as heat in the liquid film.
Throughout, the solid/liquid interface is taken as a fixed reference
plane, $z=0$ [Fig.~\ref{fig:schema}(a)]: $h$ is then simultaneously
the liquid thickness, which sets the hydrodynamic response, and the
height of the free surface, where the pulse is absorbed. This is a
modeling choice, not a physical necessity -- absorption occurs at
the (evolving) free surface while the flow is governed by the film
thickness, and the two processes are not intrinsically concomitant;
fixing the bottom is what lets a single variable $h$ carry both.
After a short thermal relaxation time -- the few picoseconds over
which electron--phonon coupling brings the lattice to local thermal
equilibrium~\cite{anisimov1974electron,hohlfeld2000electron,rethfeld2017modelling},
itself short compared with the $\sim\!100\,\mathrm{ps}$ liquid
lifetime that sets the hydrodynamic timescale below -- each liquid
column of unit base and depth $h$ therefore carries an excess areal
energy density
$Q(x,y,h) = Q_s(x,y,h) - Q_s^m$, where $Q_s$ is the energy stored
in the column and $Q_s^m$ the part consumed by melting.
We work on timescales long compared with this initial relaxation but
short enough that external losses remain small, so that to leading
order the deposited energy stays in the metal; local conversion
between liquid and solid, and loss of liquid by evaporation, are not
excluded and are represented separately below.

After thermalization the temperature is taken to be uniform within
each sufficiently narrow liquid column, so that the excess
temperature is fixed by the excess heat and by the thermal mass of
the column:
\begin{equation}
  T(x,y) - T_m
  = \frac{Q(x,y,h)}{\rho\, c_p\, h},
  \label{eq:energy_temperature}
\end{equation}
where $T_m$ is the melting temperature, and the density $\rho$ and
specific heat $c_p$ are treated as constants.
The uniform-column assumption requires the vertical diffusion time
$h_0^2/D_{\mathrm{th}}$ to be short compared with the growth time of
the instability, a condition satisfied cumulatively over successive
pulses for the films considered here; the lateral scale over which
$Q$ may be treated as uniform within a column is set by the pattern
wavelength itself and is verified a posteriori.

\subsubsection*{Thin-film equation with a source term}

In the absence of phase change, the film height obeys the
depth-integrated continuity equation
\begin{equation}
  \frac{\partial h}{\partial t} + \nabla \cdot \mathbf{q} = 0,
  \label{eq:mass_conservation}
\end{equation}
where $\mathbf{q}$ is the volumetric flux per unit width.
Within the lubrication
approximation~\cite{oron1997long,craster2009dynamics}, the
horizontal velocity obeys the Stokes balance
$\mu\,\partial_z^2\mathbf{v}_\parallel = \nabla p$, with $\mu$ the
dynamic viscosity and $p$ the pressure, no slip at the substrate,
and the tangential stress balance
$\mu\,\partial_z\mathbf{v}_\parallel|_{z=h} = \nabla\gamma$ at the
free surface: the viscous shear must balance the surface-tension
gradient, which is the driving force for thermocapillary
(Marangoni) flow~\cite{Pearson1958Sep,schatz2001experiments}.
Integrating twice and averaging over the depth gives the flux
\begin{equation}
  \mathbf{q}
  = \underbrace{-\frac{h^3}{3\mu}\nabla p}_{\displaystyle\mathbf{q}_p}
   + \underbrace{\frac{h^2}{2\mu}\nabla\gamma}_{\displaystyle\mathbf{q}_\gamma}.
  \label{eq:flux}
\end{equation}
Both the pressure-driven (Poiseuille) term $\mathbf{q}_p$ and the
stress-driven (Marangoni) term $\mathbf{q}_\gamma$ are fluxes: they
redistribute liquid laterally without changing its amount.
Phase transformations do change the local amount of liquid; we
represent them by a source term $\mathcal{S}$, whose form can be
left unspecified at this stage and is derived in
Sec.~\ref{subsec:damping}. The depth-integrated thin-film equation
is therefore
\begin{equation}
  \frac{\partial h}{\partial t}
  = \frac{1}{3\mu}\nabla\cdot\!\left(h^3\nabla p\right)
  - \frac{1}{2\mu}\nabla\cdot\!\left(h^2\nabla\gamma\right)
  + \mathcal{S}.
  \label{eq:lubrication}
\end{equation}

\subsection{Thermocapillary driving in symmetric form}
\label{subsec:marangoni}

For most metals, surface tension decreases linearly with
temperature, $\gamma(T) = \gamma_0 + \gamma'(T - T_m)$, where
$\gamma' = d\gamma/dT|_{T_m} < 0$ is the thermocapillary
coefficient. Combining with Eq.~\eqref{eq:energy_temperature} gives
the fundamental coupling between stored heat and surface tension:
\begin{equation}
  \gamma(x,y) = \gamma_0 - \Gamma\,\frac{Q(x,y)}{h(x,y)},
  \qquad
  \Gamma = \frac{|\gamma'|}{\rho\, c_p} > 0.
  \label{eq:gamma_Q}
\end{equation}
Substituting into the Marangoni term of Eq.~\eqref{eq:lubrication},
and noting that $\nabla\gamma_0 = 0$, the driving takes a remarkably
symmetric form:
\begin{align}
  -\frac{1}{2\mu}\nabla\cdot\bigl(h^2\nabla\gamma\bigr)
  &= \frac{\Gamma}{2\mu}\,\nabla\cdot\!\Bigl(h^2\,\nabla\frac{Q}{h}\Bigr)
  \nonumber\\
  &= \frac{\Gamma}{2\mu}\,\nabla\cdot\bigl(h\nabla Q - Q\nabla h\bigr)
  \nonumber\\
  &= \frac{\Gamma}{2\mu}\bigl(h\,\nabla^2 Q - Q\,\nabla^2 h\bigr).
  \label{eq:marangoni_symmetric}
\end{align}
Height and excess heat enter on the same footing: the curvature of
the heat field raises the surface with the local height as
coefficient, while the curvature of the surface lowers it with the
local heat as coefficient. Since $\Gamma > 0$, and assuming
$h, Q > 0$, the film thickens wherever
\begin{equation}
  \frac{\nabla^2 Q}{Q} > \frac{\nabla^2 h}{h}:
  \label{eq:curvature_balance}
\end{equation}
height grows where the spatial concentration of heat exceeds the
spatial concentration of the surface itself.

The pressure in Eq.~\eqref{eq:lubrication} is dominated by the
capillary (Laplace) pressure; for negligible external-pressure
gradients, $p = -\gamma_0\nabla^2 h$. No Boussinesq-type argument is
needed here: all variations of the surface tension enter through the
temperature, and the constant $\gamma_0$ therefore carries no spatial gradient,
and the variable part $\delta\gamma = \gamma(T) - \gamma_0$
contributes to the normal stress only at higher
order~\cite{oron1997long}. The thin-film equation then reads
\begin{equation}
  \partial_t h
  = -\frac{\gamma_0}{3\mu}\,\nabla\cdot\bigl(h^3\nabla\nabla^2 h\bigr)
  + \frac{\Gamma}{2\mu}\bigl(h\nabla^2 Q - Q\nabla^2 h\bigr)
  + \mathcal{S}.
  \label{eq:thin_film_full}
\end{equation}

For completeness, additional stresses may compete with or modify the
Marangoni-driven instability. Disjoining pressure becomes relevant in
nanometric films, providing either stabilizing or destabilizing
contributions depending on the sign of the van der Waals (Hamaker)
interaction, whose strength is difficult to estimate reliably for a
molten metal film on a
substrate~\cite{oron1999dewetting,ruffino2019nanostructuration}.
The vacuum recoil pressure, arising from phase change under intense
irradiation, adds a normal stress relevant in the ablation
regime~\cite{ben2007thermal}. The present analysis isolates the
thermocapillary contribution, which dominates under ultrafast,
spatially inhomogeneous heating.

\subsection{The absorption--flow feedback loop}
\label{subsec:feedback}

Before analyzing Eq.~\eqref{eq:thin_film_full} we state the feedback
loop explicitly, since every term below rests on it.
\begin{enumerate}
  \item A surface perturbation $h(x,y)$ modulates the excess heat
        stored in each liquid column, $Q(x,y,h)$, in two ways:
        through the height dependence of the optical coupling
        ($\partial_h Q < 0$) and through the thermal mass of the
        column [Eq.~\eqref{eq:energy_temperature}]. A thicker column
        stores less excess heat and dilutes it over more matter: it
        is cooler on both counts.
  \item A cooler column has a higher surface tension
        [Eq.~\eqref{eq:gamma_Q}], and the Marangoni stress drives a
        flux $\mathbf{q}_\gamma = \frac{h^2}{2\mu}\nabla\gamma$ of
        liquid toward it.
  \item The growing bump steepens the surface curvature, which
        drives the stabilizing capillary flux
        $\mathbf{q}_p = \frac{\gamma_0 h^3}{3\mu}\nabla\nabla^2 h$.
  \item The net flow alters $h(x,y)$, feeding back into step~1.
\end{enumerate}
The Marangoni step is a positive feedback: height perturbations are
amplified. The competition between this amplification and the
capillary restoring flux is what selects the wavelength $\lambda_m$
derived below.

\subsection{Linearization around a uniform film}
\label{subsec:master}

We analyze the stability of a uniform reference film by writing
$h(x,y,t) = h_0 + u(x,y,t)$ with $|u| \ll h_0$, and expand the
stored heat keeping its explicit dependence on position,
$Q(x,y,h_0+u) \approx Q_0(x,y) + Q_0'(x,y)\,u$, where
$Q_0(x,y) = Q(x,y,h_0)$ and $Q_0' = \partial_h Q|_{h_0} < 0$.
Here $Q_0$ is the heat that would be stored in the column at $(x,y)$
if its depth were the reference depth $h_0$; its spatial dependence
reflects the beam profile and any inherited topography, and is
deliberately retained; purely local models are recovered only later
as a limit.

\subsubsection*{Marangoni contribution}

Substituting into the symmetric
form~\eqref{eq:marangoni_symmetric} gives, to first order in $u$,
\begin{align*}
  h\nabla^2 Q - Q\nabla^2 h
  \approx{}& h_0\nabla^2 Q_0 + u\,\nabla^2 Q_0
           + (h_0 Q_0' - Q_0)\,\nabla^2 u \\
           &+ h_0 u\,\nabla^2 Q_0'
           + 2h_0\,\nabla Q_0'\cdot\nabla u .
\end{align*}
The combination multiplying $\nabla^2 u$ has a transparent meaning.
Define
\begin{equation}
  Q^- := Q_0 - h_0 Q_0',
  \qquad
  Q^+ := Q_0 + h_0 Q_0' :
  \label{eq:Qpm}
\end{equation}
$Q^-$ is the first-order extrapolation of the stored heat one
layer depth \emph{below} the reference surface, i.e., toward the
bottom of the film, and $Q^+$ the corresponding extrapolation one
layer depth above it [Fig.~\ref{fig:schema}(b)].
Since $Q_0' < 0$, one has $Q^+ < Q_0 < Q^-$ and $Q^- > 0$.
In terms of these quantities, the expansion assembles into the
compact form
\begin{equation}
  \left.\partial_t u\right|_M
  = \frac{\Gamma}{2\mu}\Bigl[
      h_0\nabla^2 Q_0
      + \nabla\cdot\bigl(u\,\nabla Q^+\bigr)
      - \nabla\cdot\bigl(Q^-\nabla u\bigr)
    \Bigr],
  \label{eq:marangoni_linear}
\end{equation}
the exact linear analogue of Eq.~\eqref{eq:marangoni_symmetric}.
Its two terms describe two distinct, symmetric transport channels
for the perturbation.
The second, $-\nabla\cdot(Q^-\nabla u)$, is a flux of interface
\emph{down} the gradient of $u$, with a mobility set by the heat at
the bottom of the film. Entering with a negative diffusion
coefficient, this flux acts as an anti-diffusion (liquid flows from
valleys to crests), and its strength $(\Gamma/2\mu)\,Q^-$ becomes the coefficient $\nu$
below and thereby fixes both the growth rate and the selected
wavelength. Physically, this anti-diffusion is fed from the optical
side by the same near-field enhancement of absorption in surface
troughs noted in the Introduction: nanoscale roughness locally
concentrates intensity in valleys, so the excess heat $Q^-$ that
anti-diffusion amplifies is itself seeded by that enhanced local
absorption, closing the loop between the optical and hydrodynamic
halves of the instability.
The first, $\nabla\cdot(u\,\nabla Q^+)$, is a flux of perturbation
\emph{along} the gradient of the heat extrapolated above the
surface: it transports corrugation toward the hotter regions
without amplifying it, and becomes the drift
$\mathbf{v}_{\mathrm{SH}} = (\Gamma/2\mu)\nabla Q^+$ of the exact SH form
below.
The two extrapolated heats therefore play distinct roles: $Q^-$
sets the amplification and hence $\lambda_m$, whereas $Q^+$ sets
where within the irradiated spot the pattern develops.
For a locally uniform heat profile the drift vanishes and only the
curvature balance remains: a positive perturbation grows where
$\nabla^2 Q^+/Q^- > \nabla^2 u/u$.

\subsubsection*{The reference film is not stationary}

The term $(\Gamma/2\mu)\,h_0\nabla^2 Q_0$ survives at $u = 0$: a flat film is a
stationary solution of the Marangoni dynamics only if
$\nabla^2 Q_0 = 0$ everywhere. For a bounded heat profile vanishing
far from a finite-width spot this would force $Q_0 \equiv 0$ (a
bounded harmonic function on the plane is constant), and then, since
$Q_0' < 0$, any positive perturbation would carry negative excess
heat; part of the column would be solid, a contradiction.
The flat film therefore evolves under the smooth, beam-scale forcing
$(\Gamma/2\mu)\,h_0\nabla^2 Q_0$ (supplemented below by the nonconservative
term $\mathcal{S}_0$), while the instability develops on the much
shorter pattern scale; the theory that follows describes the
perturbation $u$ about this slowly evolving reference.

\subsubsection*{Capillary contribution}

Expanding $h^3 \approx h_0^3 + 3h_0^2 u$ in the capillary flux and
keeping first-order terms gives
$\partial_t u|_c = -(\gamma_0 h_0^3/3\mu)\,\nabla^4 u$, the
standard short-wavelength stabilization.

\subsection{Evaporation, solidification, and post-pulse damping}
\label{subsec:damping}

Following the pulse, the film relaxes: electronic excitation occurs
on femtosecond timescales, but lattice heating, cooling, and surface
reorganization extend over picoseconds to
nanoseconds~\cite{hohlfeld2000electron}.
The source $\mathcal{S}$ accounts for the associated local changes
in the amount of liquid, as opposed to its lateral redistribution.
We consider two contributions,
$\mathcal{S} = \mathcal{S}_e + \mathcal{S}_s$, associated with
evaporation at the free surface and solidification at the bottom of
the liquid layer.

\subsubsection*{Evaporation}

Evaporation depends on the transient temperature of the free
surface; in the present column-averaged description, the temperature of the
reference column is $T_0 = T_m + Q_0/(\rho c_p h_0)$.
Treating evaporation as a homogeneous first-order kinetic process,
$\mathcal{S}_e = -\alpha_e(T_0)\,h$, with an Arrhenius rate
$\alpha_e \sim \nu_0\exp(-E_b/k_B T_0)$ from kinetic theory, where
$\nu_0 \sim 10^{13}\,\mathrm{s}^{-1}$ is the attempt
frequency~\cite{zangwill1988physics} and $E_b$ the binding energy
of a surface atom.
For typical parameters ($E_b \sim 0.5\,\mathrm{eV}$,
$T_0 \sim 1000\,\mathrm{K}$) one finds
$\alpha_e^{-1} \sim 30\,\mathrm{ps}$, consistent with post-pulse
cooling intervals observed in metals.
Writing $h = h_0 + u$,
$\mathcal{S}_e = -\alpha_e h_0 - \alpha_e u$: the first term thins
the reference film, the second damps its perturbations. The correction due to the increase of the actual
surface area is quadratic in $\nabla u$; it does not enter at linear
order, but it provides the key nonconservative nonlinearity in
Sec.~\ref{subsec:KS}.

\subsubsection*{Solidification}

At the bottom of the layer, heat is exchanged conductively with the
underlying solid; before solidification can proceed, the excess heat
of the liquid must be removed together with the latent heat released
by the phase change. This conductive channel is faster than cooling
through the free surface, which proceeds by radiation and
evaporation. We describe the resulting change of thickness by a
temperature-dependent rate $\mathcal{S}_s = \mathcal{S}_s(T_b)$,
with $\mathcal{S}_s < 0$ during solidification, where $T_b$ is the
temperature at the bottom of the column. Under the uniform-column assumption,
$T_b(h) = T_m + Q(h)/(\rho c_p h)$, whose derivative at $h_0$,
$h_0 Q_0' - Q_0$, is by construction exactly $-Q^-$ [the value $Q(h)$
extrapolates to at $h=0$, Fig.~\ref{fig:schema}(b)], so that to first
order $\delta T_b = -(Q^-/\rho c_p h_0^2)\,u$: the same
bottom-extrapolated heat $Q^-$ that drives the instability
controls the thermal response at the solidification front.
Linearizing $\mathcal{S}_s$ around $T_0$,
\begin{equation}
  \mathcal{S}_s(T_b) \approx \mathcal{S}_s(T_0) - \alpha_s\,u,
  \qquad
  \alpha_s = \frac{\mathcal{S}_s'(T_0)\,Q^-}{\rho c_p h_0^2}.
  \label{eq:alpha_s}
\end{equation}
A hotter liquid solidifies more slowly, giving
$\mathcal{S}_s'(T_0) > 0$; with $Q^- > 0$ this gives
$\alpha_s > 0$: a positive height perturbation is colder, solidifies
faster, and is damped.

\subsubsection*{Combined nonconservative term}

Altogether $\mathcal{S} = \mathcal{S}_0 - \alpha u$, with
\begin{equation}
  \mathcal{S}_0 = -\alpha_e h_0 + \mathcal{S}_s(T_0),
  \qquad
  \alpha = \alpha_e + \alpha_s .
  \label{eq:S_linear}
\end{equation}
$\alpha_s$ enters only through the linear response $\alpha$: it is
the derivative of $\mathcal{S}_s$ at $T_0$, not a property of
$\mathcal{S}_s(T_0)$ itself, so it plays no part in the zeroth-order
term $\mathcal{S}_0$, which retains only $\alpha_e$. The zeroth-order
term renormalizes the slow evolution of the
reference film; $-\alpha u$ damps its perturbations. The latent heat
released at the bottom reappears in the thermal budget of $Q^-$ in
Sec.~\ref{subsec:linearSH}.

\subsubsection*{Inter-pulse feedback}

The estimate $\alpha^{-1} \sim 30\,\mathrm{ps}$, material-dependent
through $\alpha_e$ and $\alpha_s$, is essential for
connecting theory with experiment. When the inter-pulse delay
satisfies $\Delta t \lesssim \alpha^{-1}$, damping is incomplete
between pulses: the surface topography from pulse $N$ is not fully
erased before pulse $N+1$ arrives. The inherited topography then
acts as a structured initial condition that biases the instability
toward specific modes. This is the physical mechanism of
\emph{morphological memory}: each pulse slightly amplifies and
refines the geometry left by its predecessor, so that the final
pattern is the cumulative result of successive pulses, each
conditioned on the previous one~\cite{banna2025photonic}. The
experimental sensitivity of morphology to $\Delta t$
(Fig.~\ref{fig:exp}) is a direct signature of this mechanism.

\subsection{Exact Swift--Hohenberg structure of the linear dynamics}
\label{subsec:linearSH}

Collecting the Marangoni, capillary, and nonconservative
contributions, the height perturbation obeys, to linear order in
$u$,
\begin{equation}
  \partial_t u
  = f
  - \nu\,\nabla^2 u
  - \beta\,\nabla^4 u
  + \mathbf{v}\cdot\nabla u
  + (r - \alpha)\,u,
  \label{eq:linear_master}
\end{equation}
with spatially varying coefficients
\begin{subequations}
\label{eq:coefficients}
\begin{align}
  f(x,y)   &= \frac{\Gamma}{2\mu}\,h_0\, \nabla^2 Q_0 + \mathcal{S}_0,
  \label{eq:forcing}\\
  \nu(x,y) &= \frac{\Gamma Q^-}{2\mu},
  \label{eq:nu}\\
  \beta    &= \frac{\gamma_0 h_0^3}{3\mu} > 0,
  \label{eq:beta}\\
  \mathbf{v}(x,y) &= \frac{\Gamma}{2\mu}\bigl(\nabla Q^+ - \nabla Q^-\bigr),
  \label{eq:drift}\\
  r(x,y)   &= \frac{\Gamma}{2\mu}\,\nabla^2 Q^+ .
  \label{eq:growth}
\end{align}
\end{subequations}
Equation~\eqref{eq:linear_master} is a variable-coefficient KS-type
equation with forcing and drift: $f$ forces the reference film
(including its nonconservative part), $\nu$ is the Marangoni
anti-diffusion coefficient, $\beta$ the capillary stabilization,
$\mathbf{v}$ a drift field, $r$ a local growth or damping rate, and
$\alpha$ the damping from evaporation and solidification.

Completing the square in the second- and fourth-order
terms (keeping in mind that the operators do not commute, since
$\nu$ is spatially varying) gives
\begin{align*}
  -\nu\nabla^2 u - \beta\nabla^4 u
  ={}& -\beta\Bigl(\nabla^2 + \frac{\nu}{2\beta}\Bigr)^{\!2} u
     + \frac{\nu^2}{4\beta}\,u \\
     &+ \nabla\nu\cdot\nabla u
     + \tfrac{1}{2}\bigl(\nabla^2\nu\bigr)\,u,
\end{align*}
so that, defining
\begin{gather}
  k_m^2 = \frac{\nu}{2\beta},
  \qquad
  \mathbf{v}_{\mathrm{SH}} = \mathbf{v} + \nabla\nu
                            = \frac{\Gamma}{2\mu}\,\nabla Q^+,
  \label{eq:km_def}\\
  \varepsilon = r - \alpha + \frac{\nu^2}{4\beta}
               + \tfrac{1}{2}\nabla^2\nu,
  \label{eq:epsilon_def}
\end{gather}
the linear dynamics takes, without further approximation, the
Swift--Hohenberg form
\begin{equation}
  \partial_t u
  = f
  + \mathbf{v}_{\mathrm{SH}}\cdot\nabla u
  + \varepsilon\,u
  - \beta\bigl(\nabla^2 + k_m^2\bigr)^2 u .
  \label{eq:SH_exact}
\end{equation}
Two physical quantities can be read off directly: $k_m$, the
wavenumber of the maximally amplified mode, and $\varepsilon$, the
local bifurcation parameter.
The SH rewriting is exact even when the coefficients vary spatially;
treating them as locally constant (on regions where $Q^\pm$, $k_m$,
and $\varepsilon$ vary little over one pattern wavelength), a mode of
wavenumber $k$ grows at the rate
\begin{equation}
  \sigma(k) = \varepsilon - \beta\bigl(k^2 - k_m^2\bigr)^2 .
  \label{eq:dispersion}
\end{equation}
The drift $\mathbf{v}_{\mathrm{SH}}$ of Eq.~\eqref{eq:SH_exact} does
not appear in $\sigma(k)$: for a locally uniform heat profile it
vanishes identically [Eq.~\eqref{eq:drift} and following], leaving
only the curvature balance that fixes $\varepsilon$ and $k_m$; it
survives only on the scale of the beam envelope, where $Q^+$ is not
locally uniform, and plays no part in the pattern-selection physics
analyzed below.

\subsubsection*{The selected wavelength and its control levers}

Growth is fastest at a single wavenumber, set by the balance between
thermocapillary destabilization and the capillary restoring stress:
\begin{equation}
  k_m^2 = \frac{3\,\Gamma Q^-}{4\,\gamma_0 h_0^3}
        = \frac{3}{4}\,s_T\,\frac{Q^-}{\rho c_p h_0^3},
  \qquad
  s_T = \frac{|\gamma'|}{\gamma_0},
  \label{eq:km_readout}
\end{equation}
where $s_T$ is the relative temperature sensitivity of the surface
tension. For $k_m$ to be real and finite we need
$Q^- = Q_0 - h_0 Q_0' > 0$, automatically satisfied since $Q_0 > 0$
and $Q_0' < 0$. The selected wavenumber increases with the
temperature sensitivity of the surface tension and with the heat
extrapolated to the bottom of the layer, and decreases with the
thermal mass of the liquid at the scale $h_0$: more stored energy
and a stronger thermocapillary response refine the pattern, whereas
a heavier, more heat-absorbing
film coarsens it.
This is also the wavenumber that maximizes the growth rate of
Eq.~\eqref{eq:dispersion}: thermocapillary anti-diffusion grows as
$\nu k^2$ while capillary stabilization grows faster, as $\beta
k^4$, so intermediate scales outgrow both the short wavelengths
capillarity suppresses and the long wavelengths anti-diffusion
amplifies too slowly to exploit. This selection is active only
while the film is liquid, over the picosecond-to-hundred-picosecond
window set by $\alpha^{-1}$ (Sec.~\ref{subsec:damping});
solidification switches it off pulse by pulse, so $k_m$ is
reselected once per pulse rather than refined continuously within
one.
Equivalently, $k_m^2 = \pi s_T\,Q^-/(m_{\mathrm{sph}}\,c_p)$, with
$m_{\mathrm{sph}} = \tfrac{4}{3}\pi\rho h_0^3$ the mass of a
liquid sphere of radius $h_0$, the thermal mass at the scale of the
film.
An equivalent form makes the scaling explicit. In terms of the
excess temperature extrapolated to the bottom,
$T^- - T_m = Q^-/(\rho c_p h_0)$, and the corresponding relative
drop of surface tension,
$\gamma^-_{\mathrm{rel}} = s_T\,(T^- - T_m)$,
\begin{equation}
  k_m^2 = \frac{3}{4}\,\frac{\gamma^-_{\mathrm{rel}}}{h_0^2}
  \quad\Longrightarrow\quad
  \lambda_m = \frac{4\pi}{\sqrt{3}}\,
              \frac{h_0}{\sqrt{\gamma^-_{\mathrm{rel}}}} :
  \label{eq:lambda_readout}
\end{equation}
the selected wavelength is proportional to the liquid depth and
inversely proportional to the square root of the relative decrease
in surface tension extrapolated to the bottom of the layer.

Table~\ref{tab:materials} collects the parameters of the two
liquid metals studied in Sec.~\ref{sec:experimental}, taken from
our earlier optical--hydrodynamic
analyses~\cite{abou2020sub,rudenko2020high}, together with the
control parameters they imply.
The reference overheating is bounded by the onset of surface
boiling ($T_0 - T_m \lesssim 1.3\times10^{3}\,\mathrm{K}$ for both
metals); we take $T_0 - T_m = 500\,\mathrm{K}$ as a representative
sub-boiling value; the bottom-extrapolated $T^-$ may exceed this
freely, since no boiling constraint applies below the surface.
Two differences between the metals stand out.
First, the thinner and less viscous Ni melt selects a finer and
faster pattern: $\lambda_m \approx 140\,\mathrm{nm}$ and
$1/\sigma_m \approx 1\,\mathrm{ns}$, against
$\approx 230\,\mathrm{nm}$ and $1.8\,\mathrm{ns}$ for
FeCr, consistent with the finer cavity arrays observed on Ni in
Fig.~\ref{fig:exp}.
Second, the two metals sit on opposite sides of the resonant zero
of the cubic coefficient derived in Sec.~\ref{subsec:SH}: for Ni,
$a = 0.77$ and the resonant $g$ is positive, and the cubic term
therefore saturates the pattern cleanly; for FeCr, $a = 1.33$ places the
film almost exactly at the zero, deferring saturation to
higher-order corrections (a contrast that echoes the clean Ni
cavity lattices against the FeCr labyrinths).
Since each pulse sustains the melt for only
${\sim}100\,\mathrm{ps}$, a few growth times require several tens
of pulses: order should build over $N \sim 10$--$30$, exactly the
range over which the Ni arrays organize in Fig.~\ref{fig:exp}.

\begin{table}[!t]
\caption{\label{tab:materials}%
Liquid-metal parameters used in this
work~\cite{abou2020sub,rudenko2020high} and derived control
parameters of the instability, for a reference overheating
$T_0 - T_m = 500$~K.
$g_{\mathrm{res}}$ is the resonant part of the cubic SH
coefficient (Sec.~\ref{subsec:SH}).}
\begin{ruledtabular}
\begin{tabular}{lcc}
 & Ni & Fe$_3$Cr \\
\hline
$\gamma_0$ (N\,m$^{-1}$) & 1.78 & 1.80 \\
$|d\gamma/dT|$ ($10^{-4}$\,N\,m$^{-1}$\,K$^{-1}$) & 4 & 4 \\
$\mu$ (mPa\,s) & 3 & 5 \\
$\rho$ (kg\,m$^{-3}$) & 7900 & 7000 \\
$c_p$ (J\,kg$^{-1}$\,K$^{-1}$) & 630 & 800 \\
$\delta$ (nm) & 13 & 15 \\
$h_0$ (nm) & 10 & 20 \\
\hline
$a = h_0/\delta$ & 0.77 & 1.33 \\
$b = \gamma^-_{\mathrm{rel}}$ & 0.29 & 0.41 \\
$\lambda_m$ (nm) & 136 & 227 \\
$1/\sigma_m$ (ns) & 1.1 & 1.8 \\
$g_{\mathrm{res}}\,h_0^2/\sigma_m$ & $+0.83$ & $-0.08$ \\
\end{tabular}
\end{ruledtabular}
\end{table}

\subsubsection*{The bifurcation parameter}

Using $r = (\Gamma/2\mu)\nabla^2 Q^+$, $\nu = (\Gamma/2\mu)\,Q^-$, and
$k_m^2 = \nu/2\beta$, the bifurcation parameter can be written
\begin{equation}
  \varepsilon
  = \frac{\Gamma}{\mu}\nabla^2 Q_0
  + \frac{\Gamma}{4\mu}\bigl(Q^- k_m^2 - \nabla^2 Q^-\bigr)
  - \alpha .
  \label{eq:epsilon_physical}
\end{equation}
The first term depends only on the curvature of the heat at the
reference depth; the second can be related to the evolution of the
heat available near the bottom of the layer, where heat is
redistributed laterally by diffusion, exchanged vertically with the
liquid above and the solid underneath, and replenished by the latent
heat released upon solidification.
Rewriting Eq.~\eqref{eq:epsilon_physical} as
\begin{equation*}
  \varepsilon
  = \frac{\Gamma}{\mu}\nabla^2 Q_0
  - \alpha_e
  - \frac{\Gamma}{4\mu}
    \Bigl[
      \nabla^2 Q^- - k_m^2 Q^- + \frac{4\alpha_s\mu}{\Gamma}
    \Bigr]
\end{equation*}
suggests identifying the bracket, up to a thermal diffusivity $D_b$
near the bottom, with the evolution of $Q^-$:
\begin{equation}
  \partial_t Q^-
  = D_b
    \Bigl[
      \nabla^2 Q^- - k_m^2\,Q^- + \frac{4\alpha_s\mu}{\Gamma}
    \Bigr].
  \label{eq:bottom_heat}
\end{equation}
The first term describes the lateral diffusion of the bottom heat;
the second, its relaxation through vertical exchange over the length
$k_m^{-1}$ selected by the hydrodynamics; the last, the latent heat
released at the liquid--solid interface.
With this identification,
\begin{equation}
  \varepsilon
  = \frac{\Gamma}{\mu}\nabla^2 Q_0
  - \alpha_e
  - \frac{\Gamma}{4\mu D_b}\,\partial_t Q^- :
  \label{eq:epsilon_bottom}
\end{equation}
the bifurcation parameter compares the curvature of the stored heat
at the reference depth with the evaporative damping and with the
rate at which heat is being removed from the bottom of the layer
($\partial_t Q^-$ measures whether heat accumulates near the bottom,
not the amount of heat itself).

\subsubsection*{Thermal versus hydrodynamic timescales}

The bottom evolution has its own timescale,
$t_b = 1/(D_b k_m^2)$, to be compared with the growth time of the
instability.
If $t_b$ is much shorter, the bottom heat reaches its stationary
balance before the surface evolves significantly:
$\partial_t Q^- \approx 0$ and
$\varepsilon \approx (\Gamma/\mu)\nabla^2 Q_0 - \alpha_e$; the marginal
condition $\varepsilon = 0$ then reads
$(\Gamma/\mu)\nabla^2 Q_0 = \alpha_e$, reducing to $\nabla^2 Q_0 = 0$ when
evaporation is negligible.
If the two timescales are comparable, the bifurcation parameter, and
with it the range of amplified wavelengths, evolves together with
the heat available at the bottom.
If $t_b$ is much longer, $Q^-$ changes little during the surface
evolution, but the bottom contribution does not vanish: the small
derivative $\partial_t Q^-$ is accompanied by the factor $1/D_b$ in
Eq.~\eqref{eq:epsilon_bottom}, and the contribution is set by the
initial departure from the stationary thermal balance.
For the materials and wavelengths considered here, the first regime
applies throughout: with a bulk metal thermal diffusivity
$D_b \sim 10^{-5}\,\mathrm{m^2/s}$ and the selected wavelengths of
Table~\ref{tab:materials} ($\lambda_m = 136\,\mathrm{nm}$ for Ni,
$227\,\mathrm{nm}$ for FeCr), $t_b = 1/(D_b k_m^2)$ is of order
$20\,\mathrm{ps}$ for Ni and $60\,\mathrm{ps}$ for FeCr --- one to
two orders of magnitude shorter than the corresponding growth times
$1/\sigma_m \approx 1.1\,\mathrm{ns}$ and $1.8\,\mathrm{ns}$. The
bottom heat therefore equilibrates essentially instantaneously on
the timescale of pattern growth, and the quasi-stationary balance
$\varepsilon \approx (\Gamma/\mu)\nabla^2 Q_0 - \alpha_e$ is the relevant one
for both materials.
Note that wherever the bottom is being cooled
($\partial_t Q^- < 0$), the last term of
Eq.~\eqref{eq:epsilon_bottom} is positive: the removal of heat from
the bottom feeds the instability and can maintain $\varepsilon > 0$
even where $\nabla^2 Q_0 < 0$, in particular at the center of a
Gaussian spot, where the quasi-stationary estimate alone would
predict damping.

\subsubsection*{Width of the amplified band}

A mode is amplified when $\sigma(k) > 0$ in
Eq.~\eqref{eq:dispersion}, which requires $\varepsilon > 0$ and
gives $|k^2 - k_m^2| < \sqrt{\varepsilon/\beta}$: the amplified
wavevectors form an annulus in wavevector space,
\begin{equation}
  k_- < |\mathbf{k}| < k_+,
  \qquad
  k_\pm = \sqrt{\,k_m^2 \pm \sqrt{\varepsilon/\beta}\,},
  \label{eq:amplified_band}
\end{equation}
centered on the critical circle $|\mathbf{k}| = k_m$.
Close to threshold, $\varepsilon \ll \beta k_m^4$ (a limit that is
generically accessible, since $k_m$ is fixed by
Eq.~\eqref{eq:km_readout} independently of $\varepsilon$), the width of the annulus is
$\delta k = k_+ - k_- \approx k_m^{-1}\sqrt{\varepsilon/\beta}
= \sqrt{4\varepsilon\mu/(\Gamma Q^-)}$: $k_m$ sets the center of the
amplified band, $\varepsilon$ its width.
If the bottom heat has reached its stationary balance and
evaporation is negligible,
$\delta k \approx 2\sqrt{\nabla^2 Q_0/Q^-}$: the width is set by the
curvature of the stored heat relative to the heat available at the
bottom.

\subsubsection*{From a KS-like to an SH-like regime}

The maximum of $\sigma(k)$ is reached at $k_m$ and equals
$\varepsilon$.
The second- and fourth-order terms of Eq.~\eqref{eq:linear_master}
contribute $\nu k^2 - \beta k^4$ to the growth rate, i.e.,
$\beta k_m^4$ at $k = k_m$; the remaining terms shift this maximum
without changing $k_m$.
Defining the effective damping
$\alpha_{\mathrm{eff}} = \alpha - r - \tfrac{1}{2}\nabla^2\nu$,
the bifurcation parameter reads
\begin{equation}
  \varepsilon = \sigma_m - \alpha_{\mathrm{eff}},
  \qquad
  \sigma_m = \beta k_m^4 = \frac{\nu^2}{4\beta}.
  \label{eq:sigma_m}
\end{equation}
Every quantity introduced from here on -- $a$, $b$, $\Lambda$,
$c_2$, $c_3$, and eventually the Swift--Hohenberg coefficients
$\gamma$ and $g$ themselves -- is a definite function of the same
nine measurable inputs already fixed by $\nu$ and $\beta$: fluence
$F$, optical skin depth $\delta$, molten-layer thickness $h_0$,
viscosity $\mu$, surface tension $\gamma_0$, its temperature
derivative $|d\gamma/dT|$, mass density $\rho$, specific heat
$c_p$, and the evaporation rate $\alpha_e$ (density and specific
heat enter only through $\Gamma = |d\gamma/dT|/\rho c_p$,
Eq.~\eqref{eq:gamma_Q}). The reader mainly interested in that
final, closed-form dependence [Eq.~\eqref{eq:coeff_brace} and
Table~\ref{tab:materials}] can take the results of the next two
subsections as established and skip ahead; they are kept here
because each abstraction -- from $\alpha_{\mathrm{eff}}$ to the
dimensionless $a,b$ -- is doing specific physical work that the
closed form alone would hide.
Two limits organize the dynamics.
When $\alpha_{\mathrm{eff}} \ll \sigma_m$, the amplified band is
broad: $\sigma(k) > 0$ for $0 < k < \sqrt{2}\,k_m$, with no lower
cutoff in an infinite domain: a KS-like instability amplifying a
wide range of scales, with $k_m$ merely the most amplified among
many.
When $\alpha_{\mathrm{eff}}$ nearly compensates $\sigma_m$,
$0 < \varepsilon \ll \sigma_m$, only a narrow annulus survives,
$\delta k/k_m \approx (1 - \alpha_{\mathrm{eff}}/\sigma_m)^{1/2}
\ll 1$: an SH-like instability.
When $\alpha_{\mathrm{eff}} > \sigma_m$, all modes are damped and
the surface stays flat.

Cooling provides a natural route through this sequence.
After the pulse, $Q^-$ decreases as the liquid cools, and the
intrinsic growth $\sigma_m = \Gamma^2 (Q^-)^2/16\mu^2\beta$ decreases with
it, approximately as $(Q^-)^2$.
The effective damping need not increase; a slower decrease than
that of $\sigma_m$ suffices for the ratio
$\alpha_{\mathrm{eff}}/\sigma_m$ to rise toward unity, narrowing the
band: from a broad KS-like regime, through a narrow SH-like window,
to complete stabilization.
This is the physical reason why ordered patterns appear only within
a finite window of fluences and inter-pulse delays.
On the pattern scale, $r$ and $\nabla^2\nu$ involve beam-scale
curvatures and are negligible, leaving $\alpha_{\mathrm{eff}}$ equal to
the damping $\alpha$ of Sec.~\ref{subsec:damping}, and the
dimensionless distance to threshold used in the remainder of the
paper is
$\tilde\epsilon = \varepsilon/\sigma_m = 1 - \alpha/\sigma_m$.

The linear analysis selects the band but not the pattern: in a
KS-like regime the nonlinearity continually redistributes energy
between wavenumbers, and the linear preference for $k_m$ does not by
itself imply spectral narrowing.
The nonlinear contributions that organize this narrowing, and then
select the symmetry of the pattern, are the object of the next
subsections.
To avoid confusion, three members of the KS/SH family appear in
this paper with distinct roles: Eq.~\eqref{eq:SH_exact} is
\emph{exact and linear}, valid for arbitrary heat profiles; the
damped KS equation~\eqref{eq:dKS} is its nonlinear completion at
small amplitude and broad spectrum in the local limit; and the
cubic SH equation~\eqref{eq:SH} is the weakly nonlinear reduction
of the same dynamics on the narrow critical band near threshold.
The last two are successive regimes, not competing models.

\subsection{Local limit: nonlinear saturation and the KS equation}
\label{subsec:KS}

On scales small compared with the beam waist, the coefficients of
Eq.~\eqref{eq:SH_exact} are locally constant: the forcing $f$ and
the drift $\mathbf{v}_{\mathrm{SH}}$ act only on the slow,
beam-scale envelope, and $Q$ depends on position only through the
local height. To describe saturation we extend the expansion of the
stored heat to third order,
\begin{equation}
  Q(h_0+u) = Q_0 + Q_0'\,u + \tfrac{1}{2}Q_0''\,u^2
           + \tfrac{1}{6}Q_0'''\,u^3 + \mathcal{O}(u^4).
  \label{eq:Q_expand}
\end{equation}

\subsubsection*{Hydrodynamic nonlinearities are conservative}

Substituting Eq.~\eqref{eq:Q_expand} into the symmetric Marangoni
form~\eqref{eq:marangoni_symmetric} and using the identities
$\nabla\cdot(u\nabla u) = \tfrac{1}{2}\nabla^2(u^2)$ and
$2u|\nabla u|^2 + u^2\nabla^2 u = \tfrac{1}{3}\nabla^2(u^3)$, all
nonlinear terms assemble into pure Laplacians:
\begin{align}
  h\nabla^2 Q - Q\nabla^2 h
  ={}& -Q^-\,\nabla^2 u
     + \frac{h_0 Q_0''}{2}\,\nabla^2(u^2) \nonumber\\
     &+ \frac{h_0 Q_0''' + Q_0''}{6}\,\nabla^2(u^3)
     + \mathcal{O}(u^4).
  \label{eq:marangoni_nonlinear}
\end{align}
This is to be expected: the Marangoni term is the divergence of a flux,
and lateral redistribution is therefore conservative to all orders; the
same holds for the capillary contributions
$-(3\beta/h_0)\,\nabla\cdot(u\nabla\nabla^2 u)$ and
$-(3\beta/h_0^2)\,\nabla\cdot(u^2\nabla\nabla^2 u)$ generated by the
expansion of $h^3$. Conservative nonlinearities redistribute liquid
between modes but cannot, by themselves, arrest the growth of the
perturbation amplitude.

\subsubsection*{The nonconservative nonlinearity}

Saturation must therefore come from the nonconservative sector.
Evaporation removes matter at a rate proportional to the true
surface area; per unit projected area,
\begin{align}
  \mathcal{S}_e
  &= -\alpha_e\,h\,\sqrt{1+|\nabla h|^2}
  \nonumber\\
  &\approx -\alpha_e h_0 - \alpha_e u - \Lambda\,|\nabla u|^2
  - \frac{\alpha_e}{2}\,u|\nabla u|^2,
  \label{eq:evap_area}
\end{align}
with $\Lambda = \alpha_e h_0/2$.
The minimal saturating nonlinearity of the KS
equation~\cite{kuramoto2003chemical,round2025coherent} thus acquires
a definite physical origin: a corrugated surface exposes more area
and evaporates faster.

\subsubsection*{The local master equation}

Collecting all terms up to third order,
\begin{align}
  \partial_t u
  ={}& -\alpha u - \nu\nabla^2 u - \beta\nabla^4 u
       - \Lambda|\nabla u|^2 \nonumber\\
     &- \frac{\alpha_e}{2}\,u|\nabla u|^2
      + \frac{\Gamma h_0 Q_0''}{4\mu}\,\nabla^2(u^2)
      \nonumber\\
     &+ \frac{\Gamma\bigl(h_0 Q_0''' + Q_0''\bigr)}{12\mu}\,\nabla^2(u^3)
      \nonumber\\
     &- \frac{3\beta}{h_0}\,\nabla\cdot(u\nabla\nabla^2 u)
      - \frac{3\beta}{h_0^2}\,\nabla\cdot(u^2\nabla\nabla^2 u).
  \label{eq:master}
\end{align}
The linear part of Eq.~\eqref{eq:master} reproduces
$k_m = \sqrt{\nu/2\beta}$ and $\sigma_m = \nu^2/4\beta$, confirming
internal consistency with Sec.~\ref{subsec:linearSH}.
Nondimensionalizing lengths by $h_0$ shows that the dynamics is
controlled by two dimensionless groups,
\begin{equation}
  a = \frac{h_0}{\delta},
  \qquad
  b = \frac{\Gamma Q^-}{\gamma_0 h_0} = \gamma^-_{\mathrm{rel}},
  \label{eq:dimless_params}
\end{equation}
Neither is a free parameter: $a$ is the purely geometric ratio of
molten-layer thickness to optical skin depth, and $b$, although
written compactly here, is fixed once $Q^-$ is -- and $Q^-$ is
itself linear in the fluence $F$ [$Q^- = F(1+2a)$ for the
exponential closure adopted below, Eq.~\eqref{eq:coeff_brace}]. $a$
enters through the absorption profile that fixes
$Q_0', Q_0'', \ldots$ (Sec.~\ref{subsec:SH}) and $b$ sets the
selected wavelength through $k_m h_0 = \sqrt{3b}/2$; both are listed
for Ni and FeCr in Table~\ref{tab:materials}, together with the
material parameters that determine them.

At small amplitude and broad spectrum, the third-order terms are
negligible and the conservative quadratic terms only redistribute
liquid; retaining the leading destabilizing, stabilizing, and
saturating terms, together with the post-pulse damping of
Sec.~\ref{subsec:damping}, gives the damped KS equation governing
the early-time dynamics:
\begin{equation}
  \partial_t u = -\alpha u - \nu\nabla^2 u - \beta\nabla^4 u
               - \Lambda|\nabla u|^2 .
  \label{eq:dKS}
\end{equation}
When $\alpha > \sigma_m$, no mode is amplified within the pulse
lifetime and the surface remains flat; the onset of patterning
requires $\alpha \lesssim \sigma_m$ (Sec.~\ref{subsec:linearSH}).

\subsection{Swift--Hohenberg equation and pattern selection}
\label{subsec:SH}

\subsubsection*{Two complementary dynamical regimes}

The KS and SH equations describe two successive and complementary
dynamical stages of the patterning process, rather than alternative
descriptions of the same physics. The conditions of validity of each
equation therefore deserve a precise statement.

The KS equation~\eqref{eq:dKS} holds when the perturbation amplitude
is small ($u \ll h_0$) but the Fourier spectrum is broad, with many
modes simultaneously active. Far from threshold, this equation
governs the early transient during which the nonlinear term
$\Lambda|\nabla u|^2$ acts as a spectral filter, suppressing modes far
from $k_m$ and progressively concentrating energy near the most
unstable wavenumber. This process does not immediately increase the
root-mean-square amplitude; the energy injected by the laser is
initially partitioned between the amplitude and the spectral width
around $k_m$. At early times the spectral width contribution
dominates, and the energy first goes into narrowing the spectrum. Only
when both contributions become comparable does the amplitude begin to
grow significantly.
This nonlinear filtering acts in the same direction as the linear
narrowing of the amplified band driven by cooling
(Sec.~\ref{subsec:linearSH}): both funnel the dynamics toward the
critical circle.

The SH equation, by contrast, is a near-threshold amplitude equation,
valid once the spectrum has already narrowed to a band of width
$\Delta k \ll k_m$ centered on $k_m$, and the effective growth rate
$\tilde\epsilon = 1 - \alpha/\sigma_m$ is small compared to unity. In
this regime, the dynamics follows the slowly varying envelope of
the pattern, and a systematic reduction becomes possible. The
transition between the two regimes occurs when the spectral width and
the perturbation amplitude become comparable:
\begin{equation}
  \frac{\Delta\lambda}{\lambda_m} \approx \frac{u}{h_0} \ll 1.
  \label{eq:transition_criterion}
\end{equation}
Typically this requires a few characteristic growth times
$1/\sigma_m$, accumulated over many successive pulses, each
sustaining the molten state for only ${\sim}100\,\mathrm{ps}$ before
resolidification.

The role of the post-pulse damping $\alpha$ in controlling which
dynamical regime is reached is illustrated in the inset of
Fig.~\ref{fig:dispersion}: when $\alpha > \sigma_m$ all modes
decay and the surface stays flat; at $\alpha = \sigma_m$ only the
single mode at $k_m$ is marginally unstable, which is the onset of
patterning ($\tilde\epsilon = 0$); for $\alpha$ slightly below
$\sigma_m$, a narrow band near $k_m$ is amplified and the SH
description applies; when $\alpha \ll \sigma_m$, a broad band is
simultaneously unstable and the dynamics is in the chaotic KS regime.

\subsubsection*{Spectral narrowing near the critical wavenumber}

\begin{figure}[t]
  \centering
  \includegraphics[width=\columnwidth]{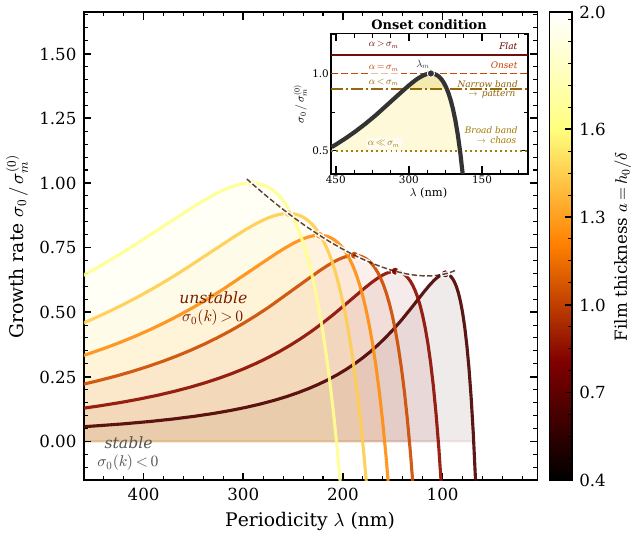}
  \caption{%
    Thermocapillary dispersion relation.
    Main panel: normalized undamped growth rate
    $\sigma_0/\sigma_m^{(0)}$ versus periodicity $\lambda$ for
    $a = h_0/\delta$ from $0.4$ (dark) to $2.0$ (light), at
    $\delta = 15\,\mathrm{nm}$ and a reference overheating
    $T_0 - T_m = 500\,\mathrm{K}$; circles mark the most-unstable
    wavelength $\lambda_m(a)$.
    Inset: for a fixed dispersion curve ($a=1.6$), the post-pulse
    damping $\alpha$ selects the dynamical regime, from flat surface
    ($\alpha > \sigma_m$) to spatiotemporal chaos
    ($\alpha \ll \sigma_m$).
  }
  \label{fig:dispersion}
\end{figure}

Setting $k = k_m(1+q)$ with $|q| \ll 1$ in the local dispersion
relation~\eqref{eq:dispersion} gives, to leading order,
\begin{equation}
  \sigma_0(k) \;\simeq\; \sigma_m\!\left[1 - 4q^2\right],
  \qquad \sigma_m = \frac{\nu^2}{4\beta},
  \label{eq:sigma_expand}
\end{equation}
where $\sigma_0(k) = \sigma(k) + \alpha$ is the undamped growth
rate: the spectrum is parabolic around its maximum at $k_m$, and the
exact operator $-\beta(\nabla^2 + k_m^2)^2$ of
Eq.~\eqref{eq:SH_exact} is precisely the differential operator with
this spectrum. In the narrow-band regime, the linear dynamics is
therefore governed by the SH operator with no further approximation.
The main panel of Fig.~\ref{fig:dispersion} shows how the
most-unstable wavelength $\lambda_m$ shifts as the ratio
$a = h_0/\delta$ increases from 0.4 to 2.0, illustrating the role of
the molten-layer thickness as the primary tuning parameter for the
selected periodicity.

\subsubsection*{The cubic SH equation near threshold}

The Swift--Hohenberg reduction of the nonlinear dynamics is not an
exact transformation of the KS equation. It arises in the
near-threshold regime, after the KS-like dynamics has already
narrowed the spectrum around $k_m$. Projecting the nonlinear terms of Eq.~\eqref{eq:master} onto the
resonant modes of the critical circle (Galerkin truncation on
$|\mathbf{k}| = k_m$), a simplification occurs: the conservative
Laplacian terms, the capillary nonlinearities, and the evaporative
slope terms all project with exactly the angular structure of $u^2$
and $u^3$ (identical self- and cross-coupling ratios), so that a
single quadratic and a single cubic coefficient suffice at this
order. The result is the cubic Swift--Hohenberg equation
\begin{equation}
  \partial_t u
  = -\alpha u - \nu\nabla^2 u - \beta\nabla^4 u    - \gamma u^2 - g u^3 ,
  \label{eq:SH}
\end{equation}
with resonant coefficients
\begin{equation}
  \gamma = k_m^2\Bigl(\frac{\Lambda}{2} + c_2\Bigr)
         + \frac{3\sigma_m}{2h_0},
  \qquad
  g = k_m^2\Bigl(c_3 + \frac{\alpha_e}{6}\Bigr)
    + \frac{\sigma_m}{h_0^2},
  \label{eq:gamma_g_res}
\end{equation}
where $c_2 = \Gamma h_0 Q_0''/4\mu$ and
$c_3 = \Gamma(h_0 Q_0''' + Q_0'')/12\mu$ are the conservative Marangoni
coefficients of Eq.~\eqref{eq:marangoni_nonlinear}, and the
$\alpha_e/6$ term arises from the third-order area correction to
evaporation.
For saturation to occur we require $g > 0$; the initial surface
roughness determines the time needed to enter the SH regime,
typically a few growth times accumulated over successive pulses.

For the exponential absorption profile
$Q(h) = F\,e^{-2(h-h_0)/\delta}$, where $F \equiv (1-R)F_{\rm inc}$
is the absorbed fluence, $R$ the reflectivity, and $\delta$ the
optical skin depth, the Taylor coefficients of
Eq.~\eqref{eq:Q_expand} read $Q_0 = F$, $Q_0' = -2F/\delta$,
$Q_0'' = 4F/\delta^2$, $Q_0''' = -8F/\delta^3$; equivalently
$q_1 = 2$, $q_2/q_1 = 1$, $q_3/q_1 = 2/3$ in the notation
$Q = Q_0(1 - q_1 u/\delta + q_2 u^2/\delta^2 - q_3 u^3/\delta^3)$.
The linear coefficients follow directly, $Q^- = F(1+2a)$ and
$\nu = \Gamma Q^-/2\mu$, with the dimensionless $a = h_0/\delta$ and
$b = \gamma^-_{\mathrm{rel}} = \Gamma Q^-/\gamma_0 h_0$ already
introduced in Eq.~\eqref{eq:dimless_params}. Together with $\beta$
[Eq.~\eqref{eq:beta}], $\sigma_m = \nu^2/4\beta$
[Eq.~\eqref{eq:sigma_m}], and the solidification damping $\alpha_s$
[Eq.~\eqref{eq:alpha_s}], this closes the chain announced there:
$\alpha$, $\nu$, $\beta$, $\gamma$, and $g$ are now fixed by the
same nine raw inputs -- fluence $F$, optical skin depth $\delta$,
molten-layer thickness $h_0$, viscosity $\mu$, surface tension
$\gamma_0$, its temperature derivative $|d\gamma/dT|$, mass density
$\rho$, specific heat $c_p$, and the evaporation rate $\alpha_e$ --
none of them adjusted to fit a pattern. The five coefficients of
the final equation of motion, Eq.~\eqref{eq:SH}, follow explicitly:
\begin{equation}
\left\{
\begin{aligned}
\nu    &= \frac{\Gamma\,Q^-}{2\mu},\\[3pt]
\beta  &= \frac{\gamma_0 h_0^3}{3\mu},\\[3pt]
\alpha &= \alpha_e + \alpha_s,\\[3pt]
\gamma &= \frac{\sigma_m}{h_0}\!\left[\frac{3}{2} + \frac{4a^2}{1+2a}\right]
            + \frac{3\,b\,\alpha_e}{16\,h_0},\\[3pt]
g      &= \frac{\sigma_m}{h_0^2}\!\left[1 + \frac{4a^2}{3}\,\frac{1-2a}{1+2a}\right]
            + \frac{b\,\alpha_e}{8\,h_0^2}.
\end{aligned}
\right.
\label{eq:coeff_brace}
\end{equation}
Table~\ref{tab:materials} gives the numerical values of $a$, $b$,
$\Gamma$, $\sigma_m$, $\alpha_s$, and of $\gamma$ and $g$
themselves for Ni and FeCr; the companion scripts carry out the
substitution symbolically as a check.
Two consequences follow.
First, $\gamma > 0$ for any stored-heat profile that is convex in
the film thickness ($Q_0'' > 0$), as is the case for exponential
absorption: the quadratic coupling then selects honeycomb voids
($\Phi$ locked to $\pi$), consistent with the circular nanocavities
observed on Ni.
Hexagonal peaks ($\gamma < 0$) require a concave stored-heat
profile ($Q_0'' < 0$), as expected for instance when absorption
saturates in thick films (a testable material criterion that
replaces the $h_0 \lessgtr \delta/4$ rule of our earlier
skin-depth closure).
Second, the resonant part of $g$ changes sign at $a \approx 1.30$
and nearly vanishes at the FeCr conditions ($a = 4/3$), leaving the
cubic saturation to the off-circle corrections.
Carrying the adiabatic elimination of the off-circle harmonics at
$q = 0$, $2k_m$, and $\sqrt{3}\,k_m$ to the next order (the
plane-wave projection is implemented in the companion scripts and
validated on the pure-SH case) restores a well-defined saturation
in the roll sector, giving for FeCr the amplitude scale
$\sqrt{g/\sigma_m} \approx 0.10\,\mathrm{nm^{-1}}$ and
$\tilde\gamma = \gamma/\sqrt{\sigma_m g} \approx 1.7$.
The hexagonal cross-coupling, by contrast, turns negative: the
voids branch is strongly subcritical, and its amplitude is limited
not by the cubic term but by the saturation of the linear response
at $|u| \sim h_0$, where the absorption expansion breaks down.
The stochastic simulations of Sec.~\ref{sec:disorder} include this
saturation explicitly.

\subsubsection*{Dimensionless form}

Rescaling $\tilde{x} = k_m x$, $\tilde{y} = k_m y$,
$\tilde{t} = \sigma_m t$, $\tilde{u} = \sqrt{g/\sigma_m}\,u$ brings
Eq.~\eqref{eq:SH} to the canonical form:
\begin{equation}
  \partial_{\tilde{t}}\tilde{u}
  = \tilde\epsilon\,\tilde{u}
  - \bigl(1 + \tilde\nabla^2\bigr)^2\tilde{u}
  - \tilde\gamma\,\tilde{u}^2
  - \tilde{u}^3,
  \label{eq:SH_dimless}
\end{equation}
with rescaled parameters:
\begin{equation}
  \tilde\epsilon = 1 - \frac{\alpha}{\sigma_m}
                 = \frac{\varepsilon}{\sigma_m},
  \qquad
  \tilde\gamma = \frac{\gamma}{\sqrt{\sigma_m g}}.
  \label{eq:dimless_params_SH}
\end{equation}
This derivation shows that the SH equation emerges directly from
thin-film hydrodynamics when nonlinear saturation in optical
absorption is retained, with all parameters tied to measurable
material and laser quantities.

\begin{figure}[!t]
  \centering
  \includegraphics[width=\columnwidth]{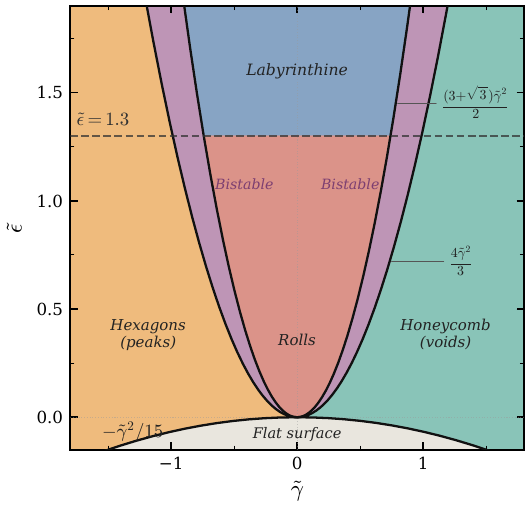}
  \caption{%
    Analytical phase diagram of the Swift--Hohenberg equation in the
    $(\tilde{\gamma},\tilde{\epsilon})$ plane.
    The boundaries $\tilde{\epsilon} = -\tilde{\gamma}^2/15$,
    $\frac{4}{3}\tilde{\gamma}^2$, and
    $\frac{3+\sqrt{3}}{2}\tilde{\gamma}^2$ separate flat surface
    (gray), hexagonal peaks (orange), honeycomb voids (teal),
    bistable coexistence (mauve), and rolls (rose); the dashed line
    marks the crossover to labyrinthine patterns (blue).
  }
  \label{fig:phase_diagram}
\end{figure}

\subsection{Pattern selection: morphological phase diagram}
\label{subsec:patterns}

\begin{figure*}[!t]
  \centering
  \includegraphics[width=\textwidth]{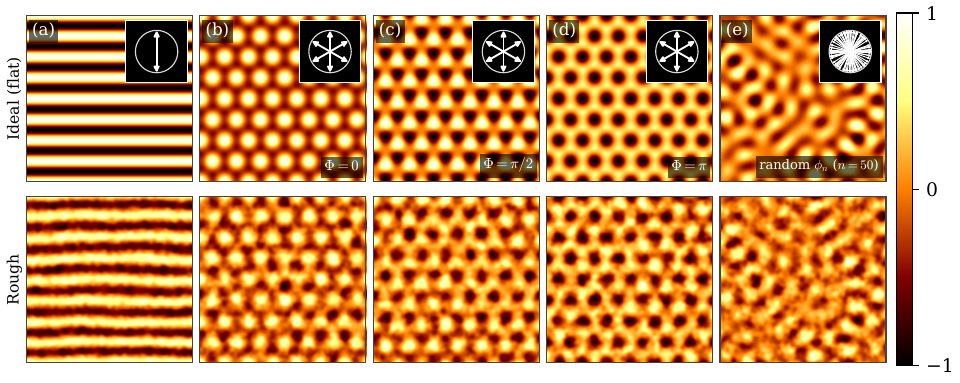}
  \caption{%
    Surface patterns assembled from plane waves on the critical
    circle $|\mathbf{k}| = k_m$ (insets: wavevector content).
    Top row (``Ideal (flat)''): the bare superposition.
    (a)~A single pair $\pm\mathbf{k}$ produces rolls.
    (b)--(d)~The same resonant triad
    ($\mathbf{k}_1+\mathbf{k}_2+\mathbf{k}_3=0$), with identical
    power spectrum, produces hexagonal peaks, an unlocked triangular
    state, or honeycomb voids depending only on the triad phase
    $\Phi = \phi_1+\phi_2+\phi_3$ [Eq.~\eqref{eq:amp_eq}].
    (e)~Many pairs ($n=50$) with random phases: an incoherent
    superposition that blends the roll, hexagonal, and honeycomb
    motifs above and can locally resemble convective cells.
    Bottom row (``Rough''): the same wavevectors and phases, with
    added phase noise and a short-range amplitude texture standing
    in for surface roughness (construction detailed in the text).
  }
  \label{fig:patterns}
\end{figure*}

\subsubsection*{Resonance conditions and the hexagonal ansatz}

The weakly nonlinear analysis of this subsection is the standard
three-mode reduction of the Swift--Hohenberg equation, established
in the pattern-formation
literature~\cite{newell1969finite,segel1969distant,cross1993pattern};
it is reproduced here because the resulting boundaries are the ones
we compare with the simulations and the experiments, and because the
coefficients entering it are the material-specific ones derived
above. No originality is claimed for the reduction itself.

The linear stage leaves a large degeneracy: it selects the modulus
of the wavevectors (the critical circle $|\mathbf{k}| = 1$), but
not their directions, and any superposition of circle modes grows
at the same rate.
The pattern symmetry is therefore decided entirely by the
nonlinearities, and the two of them act very
differently~\cite{newell1969finite,segel1969distant,newell1993order}.
The cubic term couples modes in combinations
$\mathbf{k}_1 - \mathbf{k}_1 + \mathbf{k}_2 = \mathbf{k}_2$ that
exist for any pair of directions: it saturates amplitudes but
carries no angular selectivity at leading order.
The quadratic term is the selective one.
Writing $\tilde{u} = \sum_n A_n e^{i\mathbf{k}_n\cdot\mathbf{r}}
+ \bar{A}_n e^{-i\mathbf{k}_n\cdot\mathbf{r}}$, with complex amplitude
$A_n = |A_n|\,e^{i\phi_n}$ carrying the growing magnitude $|A_n|$
and the phase $\phi_n$ of mode $n$ -- so that, written out in real
form, $\tilde{u} = \sum_n 2|A_n|\cos(\mathbf{k}_n\cdot\mathbf{r}
+\phi_n)$, with $\phi_n$ simply the offset of mode $n$'s cosine,
the same phase used for the plane waves assembled in
Fig.~\ref{fig:patterns} -- it creates from two circle modes
their sum $\mathbf{k}_n + \mathbf{k}_m$, of modulus
$|\mathbf{k}_n + \mathbf{k}_m|^2 = 2(1 + \cos\theta_{nm})$, where
$\theta_{nm}$ denotes the angle between the two directions
$\mathbf{k}_n$ and $\mathbf{k}_m$ here -- unrelated to the critical
wavenumber magnitude $k_m$ used throughout the rest of the paper.
For a generic angle $\theta_{nm}$ this offspring lies off the
circle and is damped by the linear operator: the quadratic energy
transfer is simply lost.
It survives only if the sum falls back on the circle,
$2(1 + \cos\theta_{nm}) = 1$, i.e., $\theta_{nm} = 120^\circ$: the
unique quadratic-resonant configuration is a closed triad of three
wavevector pairs at $120^\circ$,
$\mathbf{k}_1 + \mathbf{k}_2 + \mathbf{k}_3 = 0$.
This is where the three wavevectors of the hexagonal pattern come
from: they are not an ansatz but the only triplet of directions
through which the quadratic nonlinearity feeds energy coherently
back into the critical circle, the same resonance mechanism
responsible for hexagon selection in
Bénard--Marangoni convection~\cite{palm1960hexagonal,segelstuart1962preferred,golubitsky1984symmetries}.
Whenever $\tilde\gamma \neq 0$, this resonant channel amplifies
triads at second order in the amplitudes (faster, near onset, than
the cubic saturation acting at third order) and locks their phase
sum $\Phi$; rolls, built on a single pair with no triad available,
can win only where the quadratic coupling is weak.
Figure~\ref{fig:patterns} makes this mechanism visual.
A single pair of wavevectors on the critical circle produces rolls
[Fig.~\ref{fig:patterns}(a)], while one resonant triad generates
hexagonal peaks, an unlocked triangular state, or honeycomb voids
[Fig.~\ref{fig:patterns}(b--d)] depending only on the triad phase
$\Phi = \phi_1+\phi_2+\phi_3$: the three states share the same power
spectrum, and the morphology is selected by phase coherence rather
than by spectral content.
The amplitude equations derived below show that the quadratic
coupling locks $\Phi$ to $0$ or $\pi$ according to the sign of
$\tilde\gamma$, thereby selecting peaks or voids, so the unlocked
triangular state is a transient rather than an equilibrium
morphology. This near-degeneracy also explains a recurring
experimental difficulty: on a rough surface, the local six-fold
coordination that separates hexagonal peaks from the unlocked
triangular state is partly washed out by the same roughness that
broadens the phase-locking basin [Fig.~\ref{fig:patterns}, bottom
row], so the two morphologies are frequently hard to distinguish by
eye and are better classified statistically than case by case.
With many pairs and random phases [Fig.~\ref{fig:patterns}(e)], the
linear superposition already blends every motif above -- rolls,
peaks, and voids overlap locally into a texture that can resemble
convective cells -- which shows how little phase coherence survives
before the quadratic-resonance locking derived below
[Eq.~\eqref{eq:amp_eq}] imposes a single selected morphology.
The bottom row of Fig.~\ref{fig:patterns} makes the roughness
concrete: each mode's phase receives a smoothly-varying random
perturbation (Gaussian white noise, low-pass filtered in Fourier
space to a correlation length of about one pattern wavelength
$\lambda_m$, std.\ $0.35\,\mathrm{rad}$), producing the local phase
discontinuities and pattern bifurcations -- not crystallographic
defects, since the samples are single crystals -- that roughen the
lattice, and every panel additionally carries a shorter-range
amplitude texture (correlation length $\sim\!\lambda_m/10$, std.\
$12\%$ of the pattern amplitude) standing in for surface roughness.

The simplest nontrivial loop is an equilateral triangle:
$\|\mathbf{k}_1\| = \|\mathbf{k}_2\| = \|\mathbf{k}_3\| = 1$ with
$\mathbf{k}_1 + \mathbf{k}_2 + \mathbf{k}_3 = 0$. Inserting the
ansatz
$\tilde{u} = \sum_{n=1}^3 \bigl(A_n e^{i\mathbf{k}_n\cdot\mathbf{r}}
+ \bar{A}_n e^{-i\mathbf{k}_n\cdot\mathbf{r}}\bigr)$
into Eq.~\eqref{eq:SH_dimless} and projecting onto
$e^{i\mathbf{k}_1\cdot\mathbf{r}}$: the linear term
$(1+\nabla^2)^2$ vanishes since $\nabla^2 = -1$ on the unit circle.
In $\tilde{u}^2$, the only term contributing to the
$e^{i\mathbf{k}_1\cdot\mathbf{r}}$ component is
$2\bar{A}_2\bar{A}_3 e^{i(-\mathbf{k}_2 - \mathbf{k}_3)\cdot\mathbf{r}}
= 2\bar{A}_2\bar{A}_3 e^{i\mathbf{k}_1\cdot\mathbf{r}}$
since $\mathbf{k}_1 = -\mathbf{k}_2 - \mathbf{k}_3$. Collecting all
$\tilde{u}^3$ contributions proportional to
$e^{i\mathbf{k}_1\cdot\mathbf{r}}$ gives
$3(|A_1|^2 + 2|A_2|^2 + 2|A_3|^2)A_1$. The amplitude equation for
$A_1$ (and cyclically for $A_2$, $A_3$) is therefore:
\begin{equation}
  \dot{A}_1
  = \tilde\epsilon\, A_1
  - 2\tilde\gamma\,\bar{A}_2\bar{A}_3
  - 3\bigl(|A_1|^2 + 2|A_2|^2 + 2|A_3|^2\bigr)A_1.
  \label{eq:amp_eq}
\end{equation}

\subsubsection*{Hexagons and rolls as competing attractors}

Two stationary solutions of Eq.~\eqref{eq:amp_eq} compete:

\paragraph{Hexagons.}
Setting $|A_1| = |A_2| = |A_3| = A$ with equal phases gives:
\begin{equation}
  \dot{A} = \tilde\epsilon A - 2\tilde\gamma A^2 - 15A^3 = 0.
  \label{eq:hex_amp}
\end{equation}
Nontrivial stationary solutions exist when
$\tilde\gamma^2 + 15\tilde\epsilon > 0$. The sign of $\tilde\gamma$
determines the pattern type: $\tilde\gamma < 0$ yields spots
(hexagonal peaks, $\Phi$ locked to $0$), while $\tilde\gamma > 0$
yields honeycomb (voids in a hexagonal lattice, $\Phi$ locked to
$\pi$). The hexagonal state becomes unstable when
$\tilde\epsilon/\tilde\gamma^2 > (3 + \sqrt{3})/2 \approx 2.37$.

\paragraph{Rolls.}
Setting $|A_1| = A$ and $|A_2| = |A_3| \approx 0$ gives:
\begin{equation}
  \dot{A} = \tilde\epsilon A - 3A^3 = 0,
  \label{eq:roll_amp}
\end{equation}
with stable solution $A = \sqrt{\tilde\epsilon/3}$ for
$\tilde\epsilon > 0$. To assess stability of rolls against hexagonal
perturbations, one linearizes around the roll solution in the $A_2$,
$A_3$ directions:
\begin{equation}
  \dot{A}_2 = (\tilde\epsilon - 6A^2)A_2 - 2\tilde\gamma A A_3,
  \quad
  \dot{A}_3 = (\tilde\epsilon - 6A^2)A_3 - 2\tilde\gamma A A_2.
  \label{eq:roll_stability}
\end{equation}
Eigenvalue analysis shows that rolls are stable only when
$\tilde\epsilon > \frac{4}{3}\tilde\gamma^2$.

\subsubsection*{Morphological phase diagram}

The full phase diagram, summarized in Fig.~\ref{fig:phase_diagram},
is governed by the single control parameter $\tilde\epsilon/\tilde\gamma^2$:
\begin{itemize}
  \item $\tilde\epsilon/\tilde\gamma^2 < -1/15$: flat surface
        (no instability).
  \item $-1/15 < \tilde\epsilon/\tilde\gamma^2 < 4/3$: hexagons
        (peaks if $\tilde\gamma < 0$; voids if $\tilde\gamma > 0$).
  \item $4/3 < \tilde\epsilon/\tilde\gamma^2 < (3+\sqrt{3})/2$:
        bistable region; both hexagons and rolls are linearly stable,
        the selected pattern depending on initial conditions.
  \item $\tilde\epsilon/\tilde\gamma^2 > (3+\sqrt{3})/2 \approx 2.37$:
        rolls only. Under isotropic forcing, rolls align randomly
        via local imperfections, producing labyrinthine patterns.
        A phenomenological boundary at $\tilde\epsilon = 1.3$
        separates orientationally ordered rolls from the labyrinthine
        regime, reflecting the energetic penalty for defect junctions.
\end{itemize}

Since $\tilde\epsilon = 1 - \alpha/\sigma_m$ increases with fluence
$F$ and decreases with inter-pulse delay $\Delta t$ (through
$\alpha \propto e^{-\Delta t/\tau_c}$), while $\tilde\gamma$ depends
primarily on $h_0/\delta$, a model for $h_0/\delta(F)$ from
two-temperature calculations maps Fig.~\ref{fig:phase_diagram}
directly onto the experimental $(F, \Delta t)$ parameter space of
Fig.~\ref{fig:exp}, providing a physically constrained framework for
morphology selection within the reduced-model approximations.

\section{Numerical simulations}
\label{sec:numerical}

The amplitude equations of Sec.~\ref{subsec:patterns} are strictly
valid close to threshold.
To test whether their predictions survive in the fully nonlinear
regime, and to determine how sharp the morphological boundaries
actually are, we integrate Eq.~\eqref{eq:SH_dimless} numerically
over a $30\times 12$ grid of the $(\tilde{\gamma},\tilde{\epsilon})$
plane (360 simulations).
This extends the exploration initiated in
Ref.~\cite{brandao2023learning}, where SH dynamics were shown to
reproduce the sequence of laser-induced morphologies on metallic
targets, and it addresses two questions that the weakly nonlinear
analysis leaves open: whether the bistable coexistence regions
persist at finite amplitude, and where the ordered rolls give way to
labyrinths.

\subsection*{Numerical method}

Equation~\eqref{eq:SH_dimless} is integrated on a periodic square
domain of side $L = 7\times 2\pi$ (in units of $\lambda_m$),
discretized on a $128\times 128$ pseudospectral grid.
The linear operator is diagonal in Fourier space,
$\hat{\mathcal{L}}(\tilde{\mathbf{k}})
= \tilde{\epsilon} - (1 - |\tilde{\mathbf{k}}|^{2})^{2}$,
and is advanced by unconditionally stable Crank--Nicolson
half-steps, while the nonlinear terms
$-\tilde{\gamma}\tilde{u}^2 - \tilde{u}^3$ are advanced in real
space by a second-order Runge--Kutta step; the resulting Strang
splitting is second-order accurate in time
($\Delta\tilde{t} = 0.2$) and avoids the stiffness of the
fourth-order operator at short
wavelengths~\cite{cox2002exponential}.
Each run starts from a small-amplitude white-noise field, the
natural representation of a freshly polished surface whose
roughness spectrum is broad-band, and is integrated for $500$ to
$1500\,\sigma_m^{-1}$, the longer times near onset, until a
statistically stationary state is reached.

\subsection*{Probabilistic phase classification}

Assigning a single label to each converged field would
misrepresent the dynamics: the SH boundaries are crossovers rather
than sharp transitions, the outcome in the bistable region depends
on the initial condition, and at large $\tilde{\epsilon}$ the
saturated amplitude washes out the statistical signatures that
distinguish hexagons from labyrinths.
Each simulation is therefore classified probabilistically.
The analytical boundaries of Sec.~\ref{subsec:patterns} provide a
smooth prior, combined with three complementary observables
measured on the field: the kurtosis of the height distribution,
which approaches $-3/2$ for pure rolls; its skewness, which
separates spot-like from void-like patterns near onset; and the
angular entropy of the power spectrum on the critical ring, which
distinguishes orientationally ordered states from labyrinths.
The prior dominates at large $\tilde{\epsilon}$, where the field
statistics are least informative, and recedes near onset, where
they are most discriminating.
The resulting probability vectors, symmetrized under
$\tilde{\gamma}\to-\tilde{\gamma}$ (which exchanges hexagonal peaks
and honeycomb voids) and interpolated across the
$(\tilde{\gamma},\tilde{\epsilon})$ plane, yield the continuous
color field of Fig.~\ref{fig:phase_diagram_numerical}: mixed
colors flag real coexistence or sensitivity to initial
conditions, not classification failure.

\subsection*{Numerical phase diagram and morphological transitions}
\label{sec:patterns_num}

\begin{figure*}[!t]
  \centering
  \includegraphics[width=\textwidth]{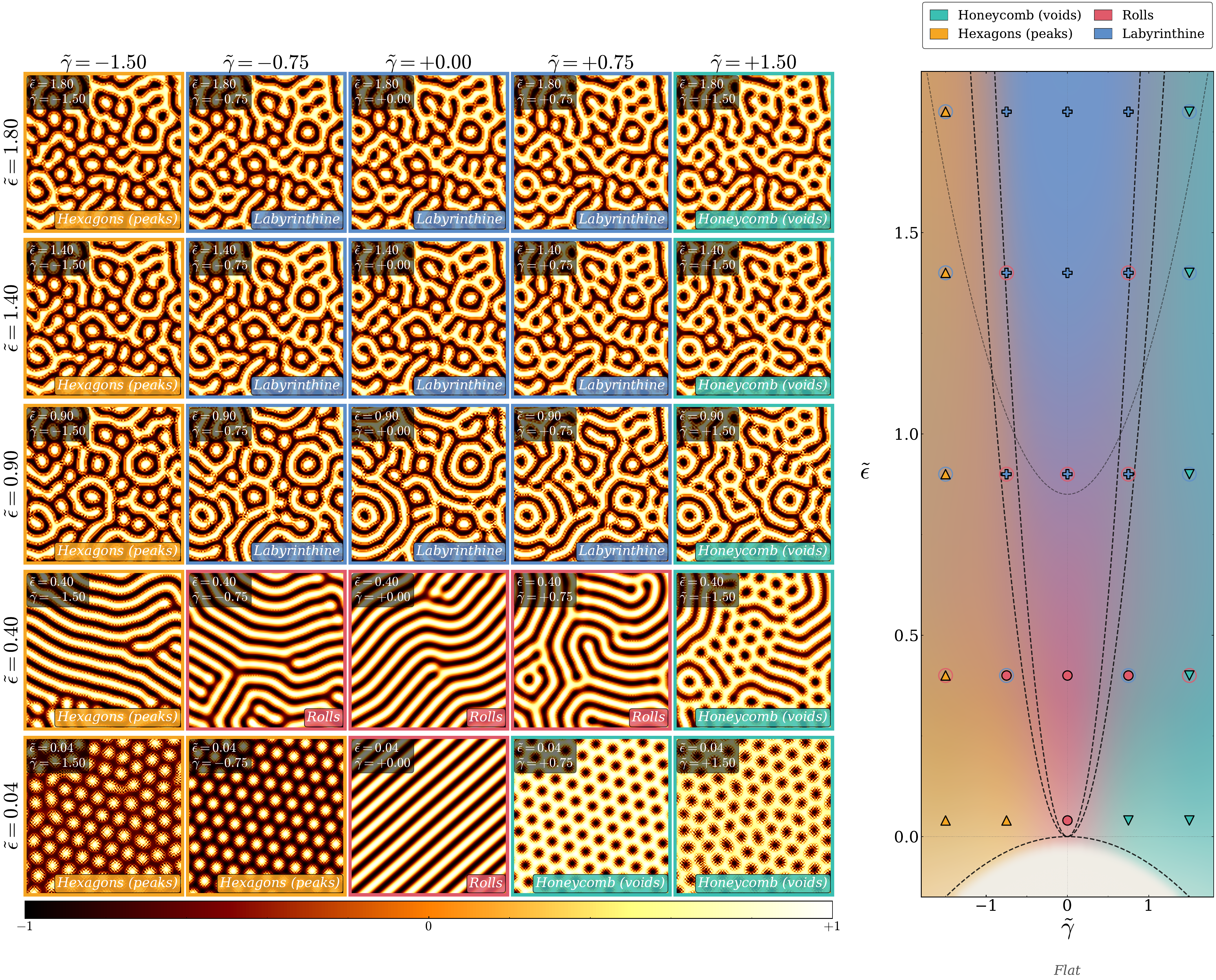}
  \caption{%
    Numerical phase diagram of the Swift--Hohenberg equation.
    Left: converged solutions of Eq.~\eqref{eq:SH_dimless} for
    $\tilde{\gamma}\in\{-1.50,\ldots,+1.50\}$ (columns) and
    $\tilde{\epsilon}\in\{0.04,\ldots,1.80\}$ (rows, bottom to top);
    border colors encode the dominant regime.
    Right: probability-weighted phase diagram built from 360
    simulations, with the analytical boundaries of
    Fig.~\ref{fig:phase_diagram} and the empirical
    rolls-to-labyrinth crossover
    $\tilde{\epsilon}=0.85+0.35\tilde{\gamma}^2$ (dashed).
    Mixed colors indicate coexistence; secondary open circles mark
    simulations where a second phase exceeds $20\%$ probability; the pale
    region below the lower boundary is the flat, subcritical surface.
  }
  \label{fig:phase_diagram_numerical}
\end{figure*}

Figure~\ref{fig:phase_diagram_numerical} confronts this numerical
map with the analytical predictions.
The boundaries derived from the three-mode amplitude equations
organize the numerical data across the whole plane, including far
from threshold where the expansion has no formal justification.
Four features of the mosaic deserve comment.

Just above onset ($\tilde{\epsilon}=0.04$, bottom row), the pattern
is decided almost entirely by the sign of the quadratic
coefficient: $\tilde{\gamma}<0$ produces hexagonal peaks,
$\tilde{\gamma}>0$ honeycomb voids, and near $\tilde{\gamma}=0$,
where no triadic resonance is available, the system falls back on
rolls [Eq.~\eqref{eq:amp_eq}].
Since the sign of $\tilde{\gamma}$ is set by the curvature of the
stored-heat profile $Q(h)$ (Sec.~\ref{subsec:SH}), the vertical
distribution of the stored heat is what decides whether the surface
develops bumps or cavities.

At moderate driving ($\tilde{\epsilon}=0.40$) the initial condition
begins to matter.
For moderate $|\tilde{\gamma}|$ both rolls and hexagons are stable
attractors~\cite{ciliberto1988competition}; starting from white
noise the rolls win, since establishing a roll requires selecting a
single wavevector whereas the hexagonal state must lock three pairs
at $60^\circ$ simultaneously~\cite{cross1993pattern}, and only a
strong quadratic coupling ($\tilde{\gamma}=\pm 1.50$) transfers
energy fast enough to lock the hexagons first.
This bistability, confirmed here at finite amplitude, implies that
the morphology reached at a given $(F,\Delta t)$ can depend on the
surface history, a further manifestation of the nonlinear
inter-pulse feedback described in Sec.~\ref{subsec:damping}.

Above $\tilde{\epsilon}\approx 0.85$ (for $\tilde{\gamma}\approx 0$)
the rolls lose orientational order: domains nucleate independently
with random orientations, no direction being preferred under
isotropic double-pulse irradiation, and the grain boundaries between
them cannot anneal, freezing the surface into a labyrinthine
texture~\cite{chate1987turbulence}.
A finite $|\tilde{\gamma}|$ delays this transition (empirically
$\tilde{\epsilon}=0.85+0.35\tilde{\gamma}^2$) by penalizing defect
junctions~\cite{cross1993pattern}.
In between, at $\tilde{\gamma}=\pm 0.75$ and
$\tilde{\epsilon}=0.90$--$1.40$, stripes coexist with hexagonal or
honeycomb patches: the numerical counterpart of the predicted
coexistence region
$\frac{4}{3}\tilde{\gamma}^2 < \tilde{\epsilon} <
\frac{3+\sqrt{3}}{2}\tilde{\gamma}^2$, which is therefore not an
artifact of the weakly nonlinear analysis.

Deep in the nonlinear regime ($\tilde{\epsilon}\geq 1.40$) every
composition turns labyrinthine, but not featureless: the network
keeps the wavelength $\lambda_m$, visible as a sharp ring in the
power spectrum, and only the long-range orientational order is lost.
This is the morphology observed on FeCr in Fig.~\ref{fig:exp}, where
the ligament networks keep a well-defined spacing without locking
into a lattice, an early indication of the coherence-limited regime
analyzed in Sec.~\ref{sec:disorder}.

The map is experimentally traceable, since $\tilde{\epsilon}$
increases with fluence and decreases with inter-pulse delay (through
$\alpha\propto e^{-\Delta t/\tau_c}$, with
$\tau_c\approx 30\,\mathrm{ps}$ for FeCr) while $\tilde{\gamma}$ is
set by $h_0/\delta$: any $(F,\Delta t)$ trajectory is a directed path
in the $(\tilde{\gamma},\tilde{\epsilon})$ plane crossing the
morphological boundaries in a predictable sequence.
On Ni, a few tens of $\mathrm{mJ\,cm^{-2}}$ separate an ordered
cavity lattice from a labyrinth of interconnected channels
(Fig.~\ref{fig:exp}), that is, a crossing of the hexagon-stability
boundary.

\subsection*{\texorpdfstring{KS\,$\to$\,SH}{KS-to-SH} spectral narrowing}
\label{sec:KS_SH}

\begin{figure*}[t]
  \centering
  \includegraphics[width=\textwidth]{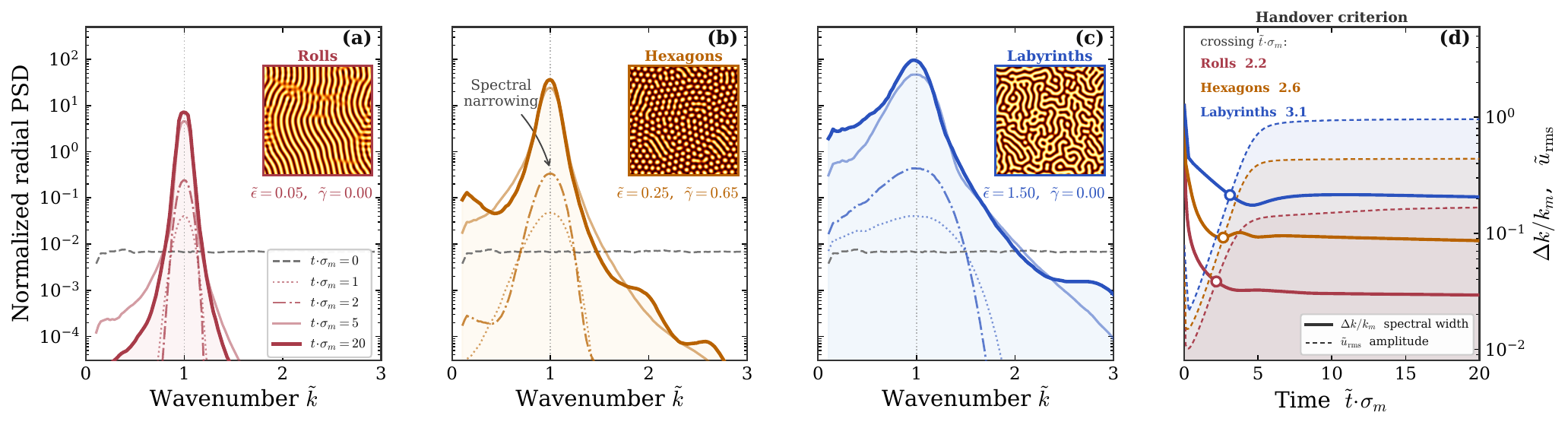}
  \caption{%
    Spectral narrowing from Kuramoto--Sivashinsky chaos to
    Swift--Hohenberg order.
    (a)--(c)~Normalized radial power spectra at
    $\tilde{t}\cdot\sigma_m = 0, 1, 2, 5, 20$ for rolls
    ($\tilde{\epsilon}=0.05$, $\tilde{\gamma}=0$), hexagons
    ($\tilde{\epsilon}=0.25$, $\tilde{\gamma}=0.65$), and labyrinths
    ($\tilde{\epsilon}=1.50$, $\tilde{\gamma}=0$), all starting from
    the same white-noise state.
    Insets: real-space fields at $\tilde{t}\cdot\sigma_m=20$.
    (d)~Test of the handover criterion~\eqref{eq:KS_SH_criterion} on
    the same runs: normalized spectral width $\Delta k/k_m$ (solid)
    and rms amplitude $\tilde{u}_\mathrm{rms}$ (dashed) versus time,
    same colors as (a)--(c). Both quantities are dimensionless, the
    amplitude being measured in units of the SH amplitude scale
    $\sqrt{\sigma_m/g}$.
    The spectrum narrows while the amplitude grows, and the open
    circles mark the crossing $\Delta k/k_m = \tilde{u}_\mathrm{rms}$
    that defines the end of the KS stage, reached after a few linear
    growth times in all three regimes.
  }
  \label{fig:ks_to_sh}
\end{figure*}
 
The handover between the two stages identified in
Sec.~\ref{subsec:SH} is set by the width of the Fourier spectrum
relative to $k_m$, and is tracked by the normalized spectral width
\begin{equation}
  \frac{\Delta k(t)}{k_m}
  = \frac{1}{k_m}
  \!\left[
    \frac{\int_0^\infty (k-k_m)^2\,\mathrm{PSD}(k,t)\,dk}
         {\int_0^\infty \mathrm{PSD}(k,t)\,dk}
  \right]^{1/2}\!\!,
  \label{eq:Delta_k}
\end{equation}
where $\mathrm{PSD}(k,t)$ is the radially averaged power spectral
density of the surface profile and $k_m=\sqrt{\nu/2\beta}$.
The handover condition (Sec.~\ref{subsec:SH}),
\begin{equation}
  \frac{\Delta k(t)}{k_m} \approx \frac{u_\mathrm{rms}(t)}{h_0},
  \label{eq:KS_SH_criterion}
\end{equation}
marks the moment when the spectral width and perturbation
amplitude are both of order $\tilde{\epsilon}^{1/2}$, at which
point the SH amplitude equation becomes valid.

Equation~\eqref{eq:KS_SH_criterion} is a prediction, and
Fig.~\ref{fig:ks_to_sh}(d) tests it directly on the same
simulations. Both sides are tracked in the dimensionless units of
Eq.~\eqref{eq:SH_dimless}, in which lengths are measured in
$k_m^{-1}$ and the surface amplitude in $\sqrt{\sigma_m/g}$, the
amplitude scale of the SH reduction; the right-hand side of
Eq.~\eqref{eq:KS_SH_criterion} then becomes the dimensionless
$\tilde{u}_\mathrm{rms}$, directly comparable with $\Delta k/k_m$.
The spectral width falls as the band narrows,
the amplitude rises as the pattern grows, and the two cross
at $\tilde{t}\cdot\sigma_m = 2.2$, $2.6$ and $3.1$ for the roll,
hexagon and labyrinth parameter sets respectively, after a few
linear growth times in every case.
Converting through $1/\sigma_m$ (Table~\ref{tab:materials}), these
crossings correspond to real handover times of
$2.4$--$3.4\,\mathrm{ns}$ for Ni and $4.0$--$5.6\,\mathrm{ns}$ for
FeCr. Measured against the single-pulse molten lifetime
(${\sim}100\,\mathrm{ps}$, Sec.~\ref{subsec:SH}), this represents on
the order of $25$--$35$ pulses for Ni and $40$--$55$ for FeCr were
the growth to proceed continuously --- an order-of-magnitude
estimate only, since in the pulsed experiments growth actually
accumulates discretely, through the inter-pulse feedback mechanism
of Sec.~\ref{subsec:damping} rather than within a single melt
duration, but one consistent with the $N\approx20$--$30$ pulse
window over which ordering is observed to develop on Ni
(Sec.~\ref{sec:experimental}).
The spectral width at the crossing follows the predicted
$\tilde{\epsilon}^{1/2}$ scaling closely,
$\Delta k/k_m = (0.18\pm 0.01)\,\tilde{\epsilon}^{1/2}$ across the
three regimes, which span a factor of $30$ in $\tilde{\epsilon}$.
Beyond the crossing the band stays narrow, $\Delta k/k_m \leq 0.21$
even in the most strongly driven case, and the narrow-band
assumption underlying the SH reduction therefore holds over the whole
interval where that reduction is used.

Panels (a) to (c) of Fig.~\ref{fig:ks_to_sh} show how the spectra
reach this state.
All three morphological regimes start from the same white-noise
initial condition ($\tilde{t}\cdot\sigma_m=0$), and undergo the
same early-time spectral narrowing. The initial spectrum ($t\cdot\sigma_m = 0$) represents the broad-band state characteristic of the KS regime, in which the liquid lifetime
per pulse (${\sim}100\,\mathrm{ps} \sim \sigma_m^{-1}$) is too short
to concentrate spectral energy around $k_m$; wavelength selection
accumulates progressively over successive pulses through morphological
memory~\cite{banna2025photonic}. The early-time evolution is controlled by the linear part of the operator alone: each Fourier mode $k\neq k_m$ is damped at rate
$|\sigma(k)|$, and within the first few linear timescales
($\tilde{t}\cdot\sigma_m=1$ and $2$) the background noise is
heavily suppressed and a sharp spectral peak emerges around
the critical wavenumber $\tilde{k}=1$.
This collapse is essentially independent of the morphological
regime: all three curves concentrate toward $\tilde{k}=1$
within $\sim\!5\,\sigma_m^{-1}$, regardless of $\tilde{\gamma}$
and the final pattern type.
The energy injected at all wavenumbers by the initial noise is
thus redirected toward $k_m$ before nonlinear saturation sets
the final amplitude.
 
The late-time spectra, by contrast, carry the explicit
signature of the final morphology.
As the system reaches its stationary state
($\tilde{t}\cdot\sigma_m=20$), the peak width of the converged
PSD encodes the symmetry of the pattern.
Rolls ($\tilde{\epsilon}=0.05$) converge to the narrowest peak,
reflecting the orientational order of a single dominant
wavevector pair.
Hexagons ($\tilde{\epsilon}=0.25$) show an intermediate width:
the three resonant mode pairs contribute at discrete azimuths
separated by $60^\circ$, adding a finite azimuthal spread to
the radially averaged PSD.
Labyrinthine patterns ($\tilde{\epsilon}=1.50$) retain the
broadest spectrum, with a persistent high-wavenumber tail arising
from the high density of grain boundaries and topological
defects that maintain a continuous distribution of roll
orientations~\cite{cross1993pattern}.

Figure~\ref{fig:ks_to_sh} thus separates the two selection events
that successive pulses sample.
The wavelength is fixed early, by the linear filter, and is common
to all morphologies; the symmetry is negotiated later by the
nonlinearities.
This is why the pattern period is robust across the experimental
conditions of Fig.~\ref{fig:exp}, while the symmetry switches
readily with fluence and delay.

\section{Disorder-limited coherence and the breakdown of long-range order}
\label{sec:disorder}
The KS--SH framework developed above rests on the assumption
that the critical hydrodynamic mode at $k_m$ remains spatially
extended throughout the irradiation sequence.
This assumption is well justified on a freshly polished surface,
where the rms roughness $\zeta_0\simeq 2\,\mathrm{nm}$ is
negligible compared to the pattern wavelength, $\lambda_m\approx
230\,\mathrm{nm}$ for the FeCr parameters used throughout this
section (Table~\ref{tab:materials}).
However, each pulse incrementally deposits corrugation, and the
cumulative topography eventually acts as a scattering potential
for subsequent pulses.
The frozen surface topography $h_N(\mathbf{x})$ left by pulse $N$
therefore enters the problem twice.
This corrugation is what pulses $1$ through $N$ have built up through
the SH instability, and at the same time the static, spatially random
modulation of the growth rate that pulse $N+1$ experiences, since the
local film thickness sets the local value of $\nu$
[Eq.~\eqref{eq:nu}].
The pattern accumulated so far is thus what scatters the mode that
continues to build it, and the natural question is whether this
self-generated scattering eventually limits the long-range
coherence, and if so at what pulse number.

\subsection*{A scattering length for the critical mode}

The frozen corrugation $h_N(\mathbf{x})$ after $N$ pulses enters
the dynamics of pulse $N+1$ through the local modulation of the
thermocapillary growth rate. From Eq.~\eqref{eq:nu},
$\nu = (\Gamma/2\mu)\,Q^-$
depends on the film thickness through the bottom-extrapolated heat,
and $h_N$ therefore shifts the local growth rate by
\begin{equation}
  \Delta\nu(\mathbf{x})
  = \frac{\partial\nu}{\partial h}\bigg|_{h_0} h_N(\mathbf{x})
  = -\,\frac{\Gamma}{2\mu}\,h_0 Q_0''\;h_N(\mathbf{x}).
  \label{eq:dnu}
\end{equation}

Since $\sigma_m = \nu^2/(4\beta)$, the maximum growth rate is
shifted by $\Delta\sigma_m/\sigma_m = 2\Delta\nu/\nu$.
In the dimensionless SH equation~\eqref{eq:SH_dimless}, where
$\tilde\epsilon = 1-\alpha/\sigma_m$, this translates into a
spatially random modulation of the control parameter:
\begin{align}
  \partial_{\tilde{t}}\tilde{u}
  &= \bigl[\tilde\epsilon + \tilde\epsilon_d(\mathbf{x})\bigr]\tilde{u}
  - \bigl(1+\tilde\nabla^2\bigr)^2\tilde{u}
  - \tilde\gamma\,\tilde{u}^2 - \tilde{u}^3,
  \nonumber\\
  \tilde\epsilon_d(\mathbf{x})
  &= -\frac{2h_0 Q_0''}{Q^-}\,h_N(\mathbf{x}),
  \label{eq:SH_disorder}
\end{align}
where $\tilde\epsilon_d$ is a zero-mean random field with the same
two-point statistics as $h_N$.
For the exponential closure at FeCr conditions,
$\tilde\epsilon_d = -[8a^2/(1+2a)]\,h_N/h_0
\approx -3.9\,h_N/h_0$: a frozen bump locally reduces the
growth rate, by thermal dilution of its column.
The sign is immaterial for the scattering statistics analyzed below,
though not for the correlation between the frozen pattern and the
growth it feeds back on.
The coherent/incoherent splitting of the disorder potential
introduced below is what makes the saturated regime tractable;
Fig.~\ref{fig:coherence_decay} below illustrates the perturbative
breakdown mechanism, while Fig.~\ref{fig:control} closes the loop
by measuring the same pulse-number dynamics directly on real
topographies.

\begin{figure}[!t]
\includegraphics[width=\columnwidth]{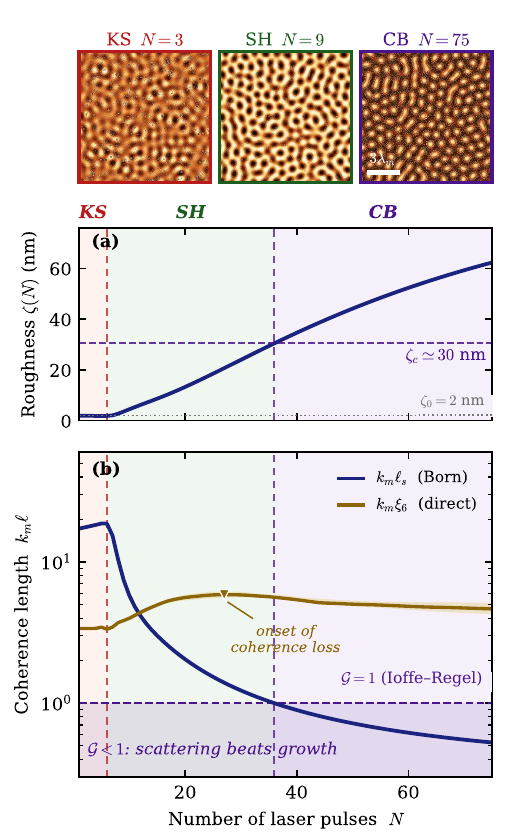}
\caption{\label{fig:coherence_decay}
Loss of pattern coherence under self-generated disorder.
Pulse-by-pulse dynamics of the stochastic Swift--Hohenberg model
[Eq.~\eqref{eq:SH_disorder}] with $\tilde\epsilon=0.20$,
$\tilde\gamma=1.00$, deposition fraction $\varphi=0.40$, and
inter-pulse relaxation $\alpha_r=2\%$; medians and interquartile
bands over four realizations, all scales fixed by the FeCr
parameters.
Top: resolidified pattern in the three successive regimes
(scale bar $3\lambda_m\approx 450\,\mathrm{nm}$).
(a)~Surface roughness $\zeta(N)$; the coherence threshold
$\zeta_c\simeq 30\,\mathrm{nm}$ is an output of the run.
(b)~Born scattering length $k_m\ell_s$ (navy), crossing the
Ioffe--Regel condition $\mathcal{G}=1$ at $N^{*}=36\pm1$, and
orientational coherence length $k_m\xi_6$ (gold), peaking at
$N\approx 27$.}
\end{figure}

\begin{figure}[!t]
\includegraphics[width=\columnwidth]{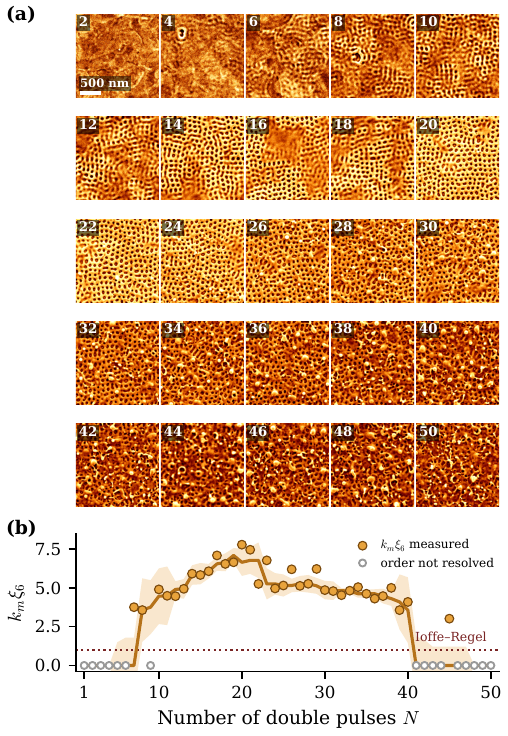}
\caption{\label{fig:control}
Orientational coherence, measured pulse by pulse, for Ni at
$230\,\mathrm{mJ/cm^2}$, $22\,\mathrm{ps}$ (same condition as
Fig.~\ref{fig:exp}).
(a)~25 of the 50 SEM frames, every other pulse from $N=2$ to $50$ in
steps of $2$ (labels on each frame; same raw images, smoothed and
recolored for display, no other processing), showing the sequence
from fine, disordered corrugation, to a well-formed hexagonal
cavity array, to a coarsened, regrown texture.
Each frame is a $2\times 2\,\mu\mathrm{m}$ detail of the full
$12.8\times 9.6\,\mu\mathrm{m}$ micrograph used for the analysis;
scale bar: 500~nm, about four pattern periods.
(b)~$k_m\xi_6(N)$ for all 50 frames (filled circles), from the
six-fold bond-orientational correlation length $\xi_6$ of the cavity
positions (Delaunay $\psi_6$, real-space correlation, exponential
fit); open circles at zero mark pulses where the near-neighbor
correlation is not statistically significant (``order not
resolved,'' $\xi_6 \equiv 0$, not omitted). Beyond $N\approx 40$ the
significance test is marginal and isolated frames may still pass it,
as at $N=45$. Solid line and shaded
band: running median and standard deviation over a $5$-pulse window,
a guide to the pulse-to-pulse scatter inherent to a single-frame
measurement, not a fit. Dotted line: the Ioffe--Regel value
$k_m\xi_6 = 1$ of Fig.~\ref{fig:coherence_decay}.}
\end{figure}

\subsubsection*{Born-approximation scattering rate}

Equation~\eqref{eq:SH_disorder} has the structure of a wave equation
with a static random potential, and the question we now ask is the
one usually asked of such equations: how far can a mode propagate
before the disorder redistributes its amplitude into other modes?
In a clean system the instability feeds a single wavevector on the
critical ring, and the pattern phase-locks over the whole
illuminated area. With $\tilde\epsilon_d\neq 0$, amplitude leaks
continuously from $\mathbf{k}_m$ into the other directions of the
ring, and the pattern can only stay coherent over the distance the
envelope travels before that leakage becomes appreciable. We
therefore need two ingredients: the rate at which amplitude leaves
$\mathbf{k}_m$, computed below in the Born approximation, and the
distance the envelope covers in that time, computed in the next
subsection. Their combination gives a coherence length, and the
condition that it exceed one pattern wavelength is the criterion we
compare with the simulations and with the measurements of
Sec.~\ref{sec:disorder}.

Expanding $\tilde{u}$ in plane waves on the critical ring
$|\tilde{\mathbf{k}}|=1$, the disorder term $\tilde\epsilon_d\tilde{u}$
couples each mode $\mathbf{k}_m$ to all other modes $\mathbf{k}'$
on the same ring via the matrix element
$V_{\mathbf{k}'\mathbf{k}_m}
 = (2/h_0)\,\widehat{h}_N(\mathbf{k}'-\mathbf{k}_m)$,
where $\widehat{h}_N$ is the Fourier transform of the frozen
corrugation. Only pairs on the ring need to be retained, because
modes off the ring are damped by the linear operator on the fast
timescale $\beta^{-1}k_m^{-4}$ and never accumulate amplitude.

Evaluating the transfer exactly would require following the mode
through repeated scattering events, each one redistributing what the
previous ones produced. The Born approximation truncates that series
at first order: the incident mode is taken to be undepleted, each
final state is reached by a single scattering event, and the total
rate is the sum of the independent probabilities $|V|^2$ over the
available final states. The approximation is the standard first step
for weak static disorder and is legitimate as long as a mode is
scattered rarely over the timescale of interest, a condition we
verify a posteriori below and which fails, by construction, at the
threshold the calculation identifies. Its value here is that it
turns the problem into a single quadrature over the measured
roughness spectrum, with no adjustable parameter.
With these assumptions the rate at which amplitude leaves
$\mathbf{k}_m$ reads
\begin{equation}
  \frac{1}{\tau_s}
  = \frac{1}{\sigma_m}
    \oint_{|k'|=k_m}
    \frac{k_m\,d\theta}{(2\pi)^2}
    \bigl\langle|V_{\mathbf{k}'\mathbf{k}_m}|^2\bigr\rangle,
  \label{eq:born}
\end{equation}
where the integral runs over scattering angle $\theta$ on the
critical ring $|k'|=k_m$, and $\sigma_m^{-1}$ is the characteristic
timescale of the SH dynamics.
Equation~\eqref{eq:born} requires only the power spectrum of the
corrugation. To obtain a closed form we model $h_N$ as a zero-mean
Gaussian random field with rms amplitude $\zeta(N)$ and lateral
correlation length $\ell_c(N)$, for which the disorder-averaged
power spectrum is
$\langle|\widehat{h}_N(\mathbf{q})|^2\rangle
 = \pi\zeta^2\ell_c^2\,e^{-q^2\ell_c^2/4}$.
This Gaussian model enters only here, in the closed-form estimates
of Eqs.~\eqref{eq:tau_s} and~\eqref{eq:mfp}; the pulse-by-pulse
simulation of Sec.~\ref{sec:disorder} evaluates the same integral on
the measured spectrum instead, and therefore does not rely on it.
For two modes on the critical ring separated by angle $\theta$,
$|\mathbf{k}'-\mathbf{k}_m|^2 = 2k_m^2(1-\cos\theta)$;
substituting and integrating over $\theta$ gives
\begin{equation}
  \frac{1}{\tau_s}
  \sim
  \frac{\zeta^2\ell_c^2 k_m}{\sigma_m h_0^2}\,
  e^{-k_m^2\ell_c^2/2},
  \label{eq:tau_s}
\end{equation}
where the Gaussian exponential arises from the power spectrum of
$h_N$ evaluated at wavevectors $q\sim k_m$.

\subsubsection*{From scattering rate to coherence length}
The second ingredient is the distance the envelope covers during
$\tau_s$. Near threshold the growth rate is parabolic about $k_m$
[Eq.~\eqref{eq:sigma_expand}], so the envelope of the critical mode
spreads diffusively rather than ballistically -- the group velocity
vanishes at the maximum of $\sigma(k)$ -- with diffusion coefficient
$D_\mathrm{env} = 2\sigma_m/k_m^2$ set by that curvature. Combining
the two ingredients, the distance over which the envelope phase
remains correlated is
$\ell_s = \sqrt{D_\mathrm{env}\,\tau_s}$, giving:
\begin{equation}
  \frac{1}{\ell_s}
  \sim
  \frac{\zeta^2\ell_c^2 k_m^2}{\sigma_m h_0^2}\,
  e^{-k_m^2\ell_c^2/2}.
  \label{eq:mfp}
\end{equation}
All prefactors involve only the KS--SH coefficients
(Eqs.~\eqref{eq:nu}--\eqref{eq:beta}), introducing no free
parameter. The exponential factor identifies a scattering
resonance: $\ell_s$ is minimized, and the disorder is therefore most
damaging, when $\ell_c\approx k_m^{-1}=\lambda_m/2\pi\approx
36\,\mathrm{nm}$. The origin of this resonance is the momentum
budget. Redirecting a mode from $\mathbf{k}_m$ to another point of
the ring requires the corrugation to supply a wavevector of modulus
$|\mathbf{k}'-\mathbf{k}_m|$, itself of order $k_m$. A corrugation
much smoother than $k_m^{-1}$ carries no Fourier weight at such
wavevectors and leaves the mode untouched, whereas a much finer one
spreads its weight over a wide range of wavevectors and supplies
little at $k_m$. The corrugation scatters most efficiently when its
own scale matches the one the pattern selects.

\subsection*{Coherence length and order degradation}
A useful dimensionless measure of pattern coherence is the ratio
$\mathcal{G}=k_m\ell_s$, which compares the coherence length with
the pattern wavelength. Its physical meaning follows from the two
timescales competing in the disordered SH dynamics: the coherent
growth time $\sigma_m^{-1}$, over which the instability builds
spatial order, and the scattering time $\tau_s$ of
Eq.~\eqref{eq:tau_s}, over which disorder redirects energy away
from $k_m$. From Eq.~\eqref{eq:mfp} and
$D_\mathrm{env}=2\sigma_m/k_m^2$ these two are related by
$\sigma_m\tau_s = \mathcal{G}^2/2$, and requiring that the
instability win the race, $\sigma_m\tau_s\gtrsim 1$, is the same as
requiring $\mathcal{G}\gtrsim 1$. The coherence length must exceed
the pattern wavelength for the SH instability to build long-range
order faster than disorder destroys it.
Scattering is therefore a source of effective non-locality in an
otherwise local field theory: energy nominally concentrated at
$k_m$ is redistributed across a neighborhood of wavevectors set by
$\ell_s$, coupling regions of the pattern separated by more than
one wavelength.

The condition $\mathcal{G}\gtrsim 1$ derived here is the analogue,
for this pattern-forming problem, of a criterion introduced by
Ioffe and Regel for electron transport in strongly disordered
solids~\cite{ioffe1960non}. Their argument is a consistency
requirement: describing a state as a propagating wave of wavevector
$k$ only makes sense if the wave survives at least one oscillation
before being scattered, that is, if its mean free path $\ell$
satisfies $k\ell \gtrsim 1$. When $k\ell$ drops to unity the wave
picture loses meaning, and in a conservative quantum system this is
where the states cease to extend across the sample and become
localized, the phenomenon identified by
Anderson~\cite{anderson1958absence,abrahams1979scaling} and later
observed for classical waves as
well~\cite{john1987strong,segev2013anderson,wiersma2013disordered}.
The same bookkeeping applies here with the critical hydrodynamic
mode in place of the electron and $\ell_s$ in place of the mean free
path, which is why $\mathcal{G}=1$ marks the point beyond which a
single extended mode no longer describes the surface.

The analogy should not be pushed further than that. Anderson
localization arises from coherent backscattering interference in a
conservative system, whereas the mechanism here is classical and
dissipative, and the modes are continuously regenerated by the
instability rather than merely propagated. What the two problems
share is precisely this non-locality: a single mode's fate is
decided by an extended disordered landscape rather than by its
immediate neighborhood. The quantity that
degrades is accordingly not a localization length but a
phase-coherence length, set by the competition between deterministic
amplification and stochastic scattering, both governed by the same
thermocapillary coefficient $\Gamma$. What the criterion provides is
a threshold, not a claim of localization.

For $\mathcal{G}\gg 1$, the scattering time greatly exceeds the
growth time, coherent pattern formation is unaffected by disorder,
and the deterministic SH phase diagram of
Fig.~\ref{fig:phase_diagram_numerical} applies without correction.
As $\mathcal{G}$ falls toward unity the pattern still nucleates
locally but can no longer phase-lock across distances larger than
$\ell_s$, and long-range orientational order degrades
progressively.

We emphasize that Eq.~\eqref{eq:mfp} is perturbative and
indicative: the Born approximation requires $\mathcal{G}\gg 1$
for self-consistency and breaks down precisely as $\mathcal{G}\to 1$
where coherence loss becomes significant. The Gaussian model for
the roughness spectrum also neglects the self-consistent evolution
of $\ell_c(N)$ under the SH instability itself. The estimate
should therefore be read as identifying the correct scaling and
the relevant physical threshold, not as a quantitative prediction
of the exact pulse number at which order collapses.

Nevertheless, the estimate is instructive for FeCr.
Using $\zeta_0\simeq 2\,\mathrm{nm}$ and
$\ell_c\ll\lambda_m/2\pi$ on the initial polished surface,
one finds $\mathcal{G}\gg 1$: the SH regime operates without
disorder correction, and the phase diagram of
Fig.~\ref{fig:phase_diagram_numerical} applies.
As corrugation grows with $N$ and $\ell_c$ drifts toward
$\lambda_m/2\pi$ under the SH instability itself, $\mathcal{G}$
decreases. For a nanostructured surface with
$\zeta\approx 25\,\mathrm{nm}$ and $\ell_c\approx k_m^{-1}$,
Eq.~\eqref{eq:mfp} yields $\mathcal{G}\approx 1$, suggesting
that coherence may be significantly degraded by this stage.
Read along the pulse sequence, this gives a definite expectation.
Over the first pulses $\mathcal{G}$ remains large, each pulse adds
corrugation without measurably shortening the coherence length, and
the deterministic phase diagram governs the outcome. As $\zeta$
accumulates, $\mathcal{G}$ decreases, and once it approaches unity
every further pulse both reinforces the pattern and scatters it,
with the second effect growing while the first saturates. Coherence
is therefore expected to degrade progressively rather than at a
sharp pulse number, for either material.
The pulse-resolved data needed to test this prediction directly are
available for Ni rather than FeCr, and the experiments there follow
the same qualitative course: order on Ni is best near
$N \approx 20$--$30$ and has visibly deteriorated by $N = 40$--$50$
(Fig.~\ref{fig:exp}), the range over which roughness accumulates
most rapidly.

\subsection*{Pulse-number evolution and process window}

Figure~\ref{fig:coherence_decay} closes the loop between the
deterministic KS--SH hierarchy and the disorder mechanism derived
above.
The model is a discrete dynamical system in the accumulated
topography $h_N$, iterated once per pulse:
$h_N \to \tilde\epsilon_d[h_N] \to u_N \to
h_{N+1} = (1-\alpha_r)h_N + \varphi\,u_N \to h_{N+1}$.
Reading the loop from left to right, the current topography defines
the disorder field through Eq.~\eqref{eq:SH_disorder}; the
Swift--Hohenberg dynamics is integrated over one melt lifetime in
that field and resolidifies into a pattern $u_N$; a fraction
$\varphi = 0.40$ of $u_N$ is frozen into the permanent topography
while a fraction $\alpha_r = 2\%$ of the existing relief relaxes
between pulses through thermal surface diffusion and partial
remelting; and the updated $h_{N+1}$ becomes the disorder seen by the
next pulse, which closes the loop.
Every scale in the simulation is fixed by the FeCr parameters used
throughout ($h_0 = 20\,\mathrm{nm}$, $\delta = 15\,\mathrm{nm}$,
$\sqrt{g/\sigma_m} = 0.185\,\mathrm{nm}^{-1}$); no quantity is
calibrated against the outcome.
The linear coupling is saturated once $|h_N|\sim h_0$, beyond which
the perturbative expansion underlying Eq.~\eqref{eq:SH_disorder} no
longer applies.
Throughout, the surface roughness $\zeta(N)$ is the spatial
root-mean-square of the resolidified topography,
$\zeta(N) = \langle h_N(\mathbf{x})^2\rangle_{\mathbf{x}}^{1/2}$,
averaged over the simulation domain and, for
Fig.~\ref{fig:coherence_decay}(a), over the four independent
disorder realizations of the run.

The run reproduces the three regimes anticipated by the analysis.
During the KS transient ($N \lesssim 6$) the spectrum narrows around
$k_m$ while the deposited amplitude remains small, and the roughness
$\zeta(N)$ [Fig.~\ref{fig:coherence_decay}(a)] stays close to the
polished-surface value $\zeta_0 = 2\,\mathrm{nm}$.
Once the triadic resonance locks the hexagonal nanocavity array,
each pulse deposits a spatially correlated corrugation faster than
inter-pulse relaxation can erase it, and $\zeta$ grows steadily
through the SH stage.

The central result is that the breakdown of order emerges from the
dynamics rather than being imposed.
The scattering length $\ell_s$ is evaluated non-parametrically from
the measured power spectrum of the accumulated disorder
[Eqs.~\eqref{eq:born}--\eqref{eq:mfp}], without assuming Gaussian
statistics for $h_N$.
The resulting coherence parameter $\mathcal{G} = k_m\ell_s$
[Fig.~\ref{fig:coherence_decay}(b), navy] decreases monotonically as
roughness accumulates and crosses the Ioffe--Regel value
$\mathcal{G} = 1$ at $N^{*} = 36 \pm 1$.
The roughness at the crossing, $\zeta_c \simeq 30\,\mathrm{nm}$, is
an output of the simulation; it agrees to within $20\%$ with the
analytic Gaussian-disorder estimate $\zeta_c \approx 25\,\mathrm{nm}$
obtained from Eq.~\eqref{eq:tau_s} with $\ell_c = k_m^{-1}$,
indicating that the Born analysis captures the correct scale even
though it formally breaks down as $\mathcal{G}\to 1$.

The crossing of $\mathcal{G}=1$ is a statement about the scattering
estimator, and confirming that it marks a loss of order, not merely
a feature of the estimator, requires a diagnostic built on a
different footing. We therefore
also measure a quantity that involves no scattering theory at all
and is read directly off the frozen topography.

The natural choice is dictated by what actually degrades. The
pattern keeps its wavelength throughout, as the sharp spectral ring
of Fig.~\ref{fig:ks_to_sh} shows, and a spectral width would be
largely blind to the effect we are after; what is lost is the mutual
alignment of the hexagonal cells across the surface. The standard
observable for exactly this kind of order is the six-fold
bond-orientational order parameter $\psi_6$ of two-dimensional
melting theory~\cite{halperin1978theory,nelson1979dislocation,zahn1999two},
defined at each cavity from the angles of the bonds joining it to
its neighbors. Its spatial correlation
$\langle\psi_6^{\vphantom{*}}(\mathbf{x})
\psi_6^{*}(\mathbf{x}+\mathbf{r})\rangle$ decays over a distance
$\xi_6$, the orientational coherence length, which is the direct
counterpart of $\ell_s$ and can be compared with it once expressed
in units of $k_m^{-1}$. When no lattice is present, the bond angles
are uncorrelated beyond nearest neighbors and $\xi_6$ collapses to
the cell spacing itself, and the diagnostic degrades gracefully
rather than returning a spurious value.
$k_m\xi_6$ [Fig.~\ref{fig:coherence_decay}(b), gold] grows through
the SH stage, peaks at $N \approx 27$, and decreases thereafter:
long-range orientational order starts to degrade as $\mathcal{G}$
approaches unity from above, before the threshold itself is
reached, the behavior expected when scattering progressively
shortens the distance over which the pattern can phase-lock.
Deep in the CB regime, $\xi_6$ settles at a few wavelengths rather
than collapsing: short-range hexagonal order survives, and the
surface resembles a structurally disordered solid that retains local
periodicity without long-range Bragg order, consistent with the
fragmented but locally ordered morphologies observed at high pulse
number in Fig.~\ref{fig:exp}.

The practical consequence is a finite process window.
High-quality nanocavity arrays require enough pulses for the SH stage to select and amplify the pattern
($N \gtrsim 6$) but not so many that self-generated roughness
destroys the coherence established in the SH stage
($N \lesssim N^{*} \approx 36$ under the present conditions).
Both bounds are experimentally accessible: the initial roughness
$\zeta_0$ is consistent with our AFM measurements of the polished
surface, and the criterion
$\zeta \lesssim \zeta_c \approx 30\,\mathrm{nm}$ provides a testable
prediction for the end of the ordering window.

\subsection*{Beyond linear response: measuring \texorpdfstring{$k_m\xi_6(N)$}{km xi6(N)} directly}

The analysis above treats $\tilde\epsilon_d$ in linear response,
and Fig.~\ref{fig:coherence_decay} operates in the corresponding
perturbative window.
The coefficients derived in Sec.~\ref{sec:nonlinear} place the
FeCr conditions beyond this window: with
$\tilde\epsilon_d \approx -3.9\,h_N/h_0$ and an amplitude scale
$\sqrt{g/\sigma_m} \approx 0.10\,\mathrm{nm^{-1}}$, the coupling
saturates within the first few pulses.
Two effects then modify the picture.
First, the potential is bounded,
$|\tilde\epsilon_d| \leq \epsilon_{\mathrm{sat}}$, and the
scattering rate, with it $1/\ell_s$, saturates as well:
roughening beyond $\zeta \sim h_0$ no longer increases the
scattering, and the linear-response estimator, which keeps growing
with $\zeta$, loses its meaning.
Second, the frozen corrugation is not generic disorder. The
accumulated topography is dominantly the pattern itself, with
spectral weight concentrated on the critical ring and phases locked
to the mode that produced it, and this becomes more pronounced as
the series progresses: the Gaussian, structureless model of
Eq.~\eqref{eq:tau_s} is at its best on the freshly polished surface
and degrades steadily with $N$, in the same direction as the
saturation just described.
Splitting the saturated potential into a component coherent with
the instantaneous pattern and an incoherent remainder,
$V = V_{\mathrm{coh}} + V_{\mathrm{dis}}$, the coherent part
scatters the critical modes into themselves (a Bragg-type pinning
that reinforces the pattern, morphological memory seen from the
scattering side), while only $V_{\mathrm{dis}}$, carried by
defects, grain boundaries, and off-ring roughness, dephases it.
The operational coherence parameter beyond linear response is
therefore $\mathcal{G}_{\mathrm{dis}} = k_m\ell_s[V_{\mathrm{dis}}]$,
evaluated non-parametrically on the saturated, pattern-orthogonal
part of the potential.

Rather than pushing the disordered-SH simulation further into this
saturated regime, we ask whether the orientational coherence
$k_m\xi_6(N)$ that it predicts can be measured directly, pulse by
pulse, on real topographies. The quantity measured is not a spectral
proxy but the same bond-orientational correlation length, computed
the same way.
We follow a single condition, Ni at $230\,\mathrm{mJ/cm^2}$,
$22\,\mathrm{ps}$, with one SEM image per pulse from $N=1$ to $N=50$
(same acquisition series as Fig.~\ref{fig:exp}). For each image, the
cavities are located, a Delaunay triangulation gives the bond
network, and the local six-fold order parameter $\psi_6$ is computed
at every site. The analysis uses the full micrograph,
$12.8\times 9.6\,\mu\mathrm{m}$, which contains between
$6\times 10^{3}$ and $8.5\times 10^{3}$ cavities; panel~(a) of
Fig.~\ref{fig:control} reproduces a $2\times 2\,\mu\mathrm{m}$ detail
of the same frames, enlarged so that individual cavities stay
visible.
The measured density, $52\,\mu\mathrm{m}^{-2}$ at $N=45$, gives a
mean spacing of $139\,\mathrm{nm}$, consistent with the
$\lambda_m = 136\,\mathrm{nm}$ predicted for Ni in
Table~\ref{tab:materials}. With that many sites the correlation
function is averaged over more than $10^{7}$ pairs, and the
statistical uncertainty on $\xi_6$ is dominated by frame-to-frame
variability rather than by counting noise.
The real-space correlation $\langle\psi_6(0)\psi_6^{*}(r)\rangle$,
averaged over all pairs and binned in $r$, decays exponentially, and
$\xi_6(N)$ is the decay length of a single-parameter exponential fit
to that profile. Apart from that decay length and the pixel
calibration of the micrograph, the procedure introduces no adjustable
quantity: the cavity positions, the bond network and $\psi_6$ itself
all follow from the image. Where the near-neighbor correlation is not
statistically significant against the large-$r$ noise floor, meaning
that no orientational order is resolved, $\xi_6$ is set to zero
rather than reported from an ill-defined fit.

Panel (a) of Fig.~\ref{fig:control} shows the
KS $\to$ SH $\to$ decoherence sequence directly: no resolved order
over the first several pulses, a well-formed hexagonal array between
$N \approx 12$ and $N \approx 30$, and a progressively regrown,
finer-grained texture beyond $N \approx 32$. Panel (b) quantifies
it: $k_m\xi_6$ is indistinguishable from zero for
$N \lesssim 6$, rises sharply, peaks around $N \approx 20$ at
$k_m\xi_6 \approx 7$--$8$, then declines, falling back to an
ill-defined, sub-Ioffe--Regel signal by $N \approx 41$--$50$. This
reproduces, pulse for pulse, the rise-then-decline shape that the
stochastic SH simulation of Fig.~\ref{fig:coherence_decay}(b)
predicts for $k_m\xi_6(N)$, over the same order of magnitude in $N$:
the simulation, run at the generic FeCr parameters used throughout,
peaks near $N \approx 27$ and crosses the Ioffe--Regel threshold at
$N^{*} = 36 \pm 1$; the measurement, on an independent condition and
material, peaks a few pulses earlier and has largely collapsed by
$N \approx 40$. Neither the pulse at which a real surface first
nucleates a resolvable lattice nor the rate at which self-generated
roughness accumulates is an input to the simulation, and the data
played no part in fixing any coefficient in
Secs.~\ref{sec:nonlinear}--\ref{sec:disorder}. The agreement is
one of mechanism, shape and order of magnitude in $N$ rather than of
individual pulse numbers.

\section{Conclusion}
\label{sec:conclusion}

Starting from the lubrication form of the Navier--Stokes equations,
we have derived a hierarchy of reduced nonlinear models for
ultrafast laser nanopatterning on metallic surfaces, with every
coefficient traced to measurable laser and material parameters
rather than fitted to the observed morphology.
The KS and SH coefficients are expressed in terms of the laser
fluence $F$, optical skin depth $\delta$, thermocapillary
coefficient $\Gamma$, and mean molten-layer thickness $h_0$,
closing the chain from single-pulse hydrodynamics to the pattern
deposited after many pulses.

The thermocapillary instability is driven by a self-reinforcing
feedback loop in which a surface corrugation modulates local
absorption, creating a surface-tension gradient that drives
Marangoni flow and deepens the corrugation further.
Competition between this destabilizing flux and the capillary
restoring stress selects a wavelength set by the melt depth and
skin depth alone, finer on Ni ($\lambda_m\approx136\,\mathrm{nm}$)
than on FeCr ($\lambda_m\approx227\,\mathrm{nm}$,
Table~\ref{tab:materials}).
Post-pulse relaxation introduces a material-dependent linear damping
$\alpha$ ($\sim30\,\mathrm{ps}$ for FeCr, through the evaporation and
solidification rates $\alpha_e$ and $\alpha_s$) that governs
inter-pulse feedback: when $\Delta t\lesssim\alpha^{-1}$, the
topography left by pulse $N$ acts as a structured seed for pulse
$N+1$, biasing the instability toward progressively ordered states.

Pattern formation proceeds through three successive dynamical
regimes.
In the \textit{KS regime}, linear growth-rate selectivity
concentrates spectral energy toward $k_m$: each mode away from
$k_m$ decays at its own rate $|\sigma(k)|$, so noise collapses onto
a narrow ring within a few linear growth times, while the
broad-spectrum Marangoni nonlinearity $\Lambda|\nabla u|^2$, active
because many modes are still comparably excited, only caps the
growing amplitude without locking a symmetry; the surface therefore
stays small in amplitude and no lattice is resolved. Both the
simulation and the pulse-by-pulse
measurement place the end of this stage near the sixth pulse,
$k_m\xi_6$ being indistinguishable from zero for $N\lesssim 6$
(Fig.~\ref{fig:control}).
In the \textit{SH regime}, the narrow-band amplitude equation
selects the pattern symmetry via resonant triadic interactions:
the single ratio $\tilde{\epsilon}/\tilde{\gamma}^2$ governs
transitions between flat surface, hexagonal nanocavity arrays,
bistable domains, and labyrinthine rolls, in quantitative
agreement with the numerical phase diagram of
Fig.~\ref{fig:phase_diagram_numerical}.
This ordered stage is self-limiting: the corrugation frozen in at
each resolidification is both what the instability has built and what
randomly modulates the growth rate seen by the next pulse.
In the stochastic SH simulation, with every scale fixed by the
FeCr parameters, the coherence parameter
$\mathcal{G} = k_m\ell_s$ crosses the Ioffe--Regel condition
$\mathcal{G} = 1$ at $N^{*} = 36\pm 1$, at an accumulated
roughness $\zeta_c \simeq 30\,\mathrm{nm}$ close to the analytic
Gaussian-disorder estimate of $25\,\mathrm{nm}$
(Eq.~\eqref{eq:mfp}); the orientational coherence length $\xi_6$
peaks slightly earlier ($N\approx 27$) and decreases as the
threshold is approached.
The resulting morphologies (aperiodic bumps and disordered
networks retaining local hexagonal packing) correspond to a
disorder-limited regime in which no extended hydrodynamic
eigenmode dominates.
The Born estimate is perturbative and breaks down precisely at
the threshold it identifies; these numbers locate the crossover
rather than predict it sharply. Beyond linear response, the
saturation of the disorder coupling protects the pattern that the
instability generates, the generalized scattering estimator remains
above the Ioffe--Regel threshold, and both agree with the directly
measured orientational coherence.

Pulse-resolved experiments on Ni and FeCr support this hierarchy.
The selected scale appears early and stays fixed while the
symmetry develops, and is finer on Ni than on FeCr, as the
derived coefficients require (Table~\ref{tab:materials}).
The two metals fall on opposite sides of the resonant zero of the
cubic coefficient, and correspondingly one orders into cavity
lattices while the other remains labyrinthine.
On Ni, the degree of order is non-monotonic in the pulse number.
This behavior is qualitatively visible across the series of
Fig.~\ref{fig:exp} and is quantified by the bond-orientational
measurement of Fig.~\ref{fig:control}: computed on the SEM frames
with the cavity positions as its only input, $k_m\xi_6(N)$ rises out
of the KS regime, peaks at $N\approx 20$, and decays back to an
ill-defined signal by $N\approx 41$--$50$.
The ordering window is therefore finite, bounded below by pattern
selection and above by disorder-driven coherence loss, exactly as
the stochastic SH simulation predicts, and here measured on real
topographies rather than simulated ones.

Several extensions follow naturally.
For linearly polarized irradiation, the isotropic
$|\nabla u|^2$ nonlinearity must be replaced by a tensorial
form that breaks rotational symmetry and selects oriented
LIPSS~\cite{zhang2020laser,perrakis2024impact}.
The disorder feedback treated here perturbatively calls for a
self-consistent description beyond the Born approximation, one that
retains the non-Gaussian statistics and the evolving correlation
length of the accumulated corrugation.
On the experimental side, Fig.~\ref{fig:control} tests the predicted
threshold through the orientational coherence alone; a sharper test
is within reach, since the theory predicts a critical roughness
$\zeta_c \approx 30\,\mathrm{nm}$, and measuring $\zeta(N)$ by
pulse-resolved AFM on the same surfaces, for several fluences and
delays, would confront that number directly instead of inferring it
from $\xi_6$.
A further, more speculative direction follows from how strongly
$\lambda_m$ and the pattern symmetry depend on $h_0/\delta$ and
$\Gamma$: a spatially modulated thermocapillary coefficient
$\Gamma(x,y)$, achievable in principle through compositional
patterning or surface functionalization, would let the local
instability wavelength and pattern symmetry be
programmed by material design rather than by beam shaping, a
possible route to lithography-free fabrication of prescribed
nanoarchitectures.

Taken together, these results show that a molten metal surface
driven far from equilibrium does not merely relax: pulse after
pulse, it converts a structureless perturbation into a reproducible,
material-encoded architecture, and eventually into its own disorder.
The nonlinear hydrodynamics developed here turns that
self-organization from a qualitative analogy into a quantitative,
predictive theory, one that ties the pattern deposited after many
pulses to laser and material parameters fixed before the first pulse
ever arrives.

\begin{acknowledgments}
This work was supported by the project MELISSA (ANR-24-CE23-7140-01) funded by the French Agence Nationale de la
Recherche.\end{acknowledgments}

\bibliographystyle{apsrev4-2}
\bibliography{References}

\end{document}